\documentclass[10pt,letterpaper]{article}

\usepackage{jheppub_mod}
\usepackage{slashed}
\usepackage{feynmp}

\newcommand{\ga}{\gamma}
\newcommand{\ep}{\epsilon}
\newcommand{\ka}{\kappa}
\newcommand{\lam}{\lambda}
\newcommand{\sig}{\sigma}

\newcommand{\Ga}{\Gamma}
\newcommand{\La}{\Lambda}

\newcommand{\ve}{\varepsilon}
\newcommand{\vt}{\vartheta}

\newcommand{\cD}{\mathcal{D}}
\newcommand{\cF}{\mathcal{F}}
\newcommand{\cG}{\mathcal{G}}
\newcommand{\cL}{\mathcal{L}}
\newcommand{\cK}{\mathcal{K}}
\newcommand{\cM}{\mathcal{M}}
\newcommand{\cN}{\mathcal{N}}
\newcommand{\cO}{\mathcal{O}}

\newcommand{\ud}{\mathrm{d}}
\newcommand{\ue}{\mathrm{e}}
\newcommand{\ui}{\mathrm{i}}
\newcommand{\us}{\mathrm{s}}

\newcommand{\uD}{\mathrm{D}}
\newcommand{\uL}{\mathrm{L}}
\newcommand{\uR}{\mathrm{R}}

\newcommand{\vp}{\varphi}

\newcommand{\II}{\mathbb{I}}

\newcommand{\vol}{\mathrm{vol}}
\newcommand{\Str}{\mathrm{Str}}
\newcommand{\tr}{\mathrm{tr}}

\newcommand{\SO}[1]{\mathrm{SO}\left(#1\right)}
\newcommand{\SU}[1]{\mathrm{SU}\left(#1\right)}
\newcommand{\U}[1]{\mathrm{U}\left(#1\right)}

\newcommand{\ul}[1]{\underline{#1}}
\newcommand{\bs}[1]{\boldsymbol{#1}}

\makeatletter
\newsavebox\myboxA
\newsavebox\myboxB
\newlength\mylenA

\newcommand*\xoverline[2][0.75]{%
    \sbox{\myboxA}{$\m@th#2$}%
    \setbox\myboxB\null
    \ht\myboxB=\ht\myboxA%
    \dp\myboxB=\dp\myboxA%
    \wd\myboxB=#1\wd\myboxA
    \sbox\myboxB{$\m@th\overline{\copy\myboxB}$}
    \setlength\mylenA{\the\wd\myboxA}
    \addtolength\mylenA{-\the\wd\myboxB}%
    \ifdim\wd\myboxB<\wd\myboxA%
       \rlap{\hskip 0.5\mylenA\usebox\myboxB}{\usebox\myboxA}%
    \else
        \hskip -0.5\mylenA\rlap{\usebox\myboxA}{\hskip 0.5\mylenA\usebox\myboxB}%
    \fi}
\makeatother

\title{Soft branes in supersymmetry-breaking backgrounds}

\date{\today}

\author[a]{Paul McGuirk,}
\author[b,c,d]{Gary Shiu,}
\author[b]{and Fang Ye}

\affiliation[a]{Laboratory for Elementary-Particle Physics, Cornell
  University, Ithaca, New York, USA}
\affiliation[b]{Department of
  Physics, University of Wisconsin-Madison, Madison, Wisconsin, USA}
\affiliation[c]{Institute for Advanced Study, Hong Kong University of
  Science and Technology, Hong Kong}
\affiliation[d]{Institute for
  Theoretical Physics, University of Amsterdam, Amsterdam, the
  Netherlands}

\emailAdd{mcguirk@cornell.edu}
\emailAdd{shiu@physics.wisc.edu}
\emailAdd{fye6@wisc.edu}

\abstract{We revisit the analysis of effective field theories
  resulting from non-supersymmetric perturbations to supersymmetric
  flux compactifications of the type-IIB superstring with an eye
  towards those resulting from the backreaction of a small number of
  $\overline{\uD 3}$-branes.  Independently of the background, we show
  that the low-energy Lagrangian describing the fluctuations of a
  stack of probe D3-branes exhibits soft supersymmetry breaking,
  despite perturbations to marginal operators that were not fully
  considered in some previous treatments.  We take this as an
  indication that the breaking of supersymmetry by $\overline{\uD
    3}$-branes or other sources may be spontaneous rather than
  explicit.  In support of this, we consider the action of an
  $\overline{\uD 3}$-brane probing an otherwise supersymmetric
  configuration and identify a candidate for the corresponding
  goldstino.}

\preprint{MAD-TH-12-04}

\begin{document}

\maketitle

\section{Introduction}

A persistent problem in the development of realistic string
compactifications is the implementation of supersymmetry breaking in a
genuinely stringy and controllable manner.  The tension comes from the
fact that string theory is usually defined in 10 (or 11) dimensions
with a large number of supercharges, while realistic phenomenology
requires such compactifications to reduce at low energies to
non-supersymmetric 4d theories.  Added to this tension is the issue of
moduli stabilization whose details can significantly affect the vacuum
structure and supersymmetry-breaking terms in the dimensionally
reduced theories.  One can contemplate bypassing the intermediate
stage of realizing an effective 4d supergravity at low energies by
constructing stabilized vacua with supersymmetry broken at or above
the compactification scale.  The construction of such vacua has proven
to be a difficult task as one often encounters (perturbative)
instabilities.  Thus far, explicit tachyon-free examples of this kind
with the broad features of the Standard Model have not yet been found,
and there are generic statistical arguments suggesting that such vacua
are rare~\cite{Chen:2011ac} (see also~\cite{Marsh:2011aa}).
Furthermore, the scale of supersymmetry breaking in such constructions
is typically much larger than the electroweak scale $m_{Z}\sim 100\
\mathrm{GeV}$, with no apparent relation between them.  In light of
these issues, studies of supersymmetry breaking in string theory often
takes a different route.  Most work on the subject begins with an
effective 4d supergravity, as there are several potential
phenomenological benefits for supersymmetry (at least the reduced
version, e.g., $\cN_{4}=1$) to persist at intermediate scales.  Other
than protecting certain operators from large quantum corrections,
subsequent breaking of supersymmetry in the effective low-energy
supergravity provides a nice mechanism to trigger spontaneous
electroweak symmetry breaking thus tying the electroweak scale to the
supersymmetry-breaking scale.  Furthermore, such a framework of
intermediate or low-scale breaking has the advantage that a myriad of
$\cN_4=1$ string constructions with semi-realistic spectra are readily
available, while there are fewer examples for those exhibiting
high-scale supersymmetry breaking. Traditionally, the source of
supersymmetry breaking in this context is assumed to be the result of
some hidden sector dynamics. Recent developments have extended the
possibilities to include other supersymmetry-breaking sources such as
fluxes and anti-branes.  It is useful to note that while we make a
distinction for those constructions that admit a 4d supergravity
description at an intermediate scale, such vacua, when lifted to 10d,
should correspond to a non-supersymmetric background when the
backreaction of the supersymmetry-breaking effects is taken into
account\footnote{The supersymmetry-breaking effects here are not
  restricted to localized sources, but include also fluxes as well as
  sources of dynamical supersymmetry breaking in the hidden sector
  realized as instantons on branes.}. Thus, from a 10d point of view,
the nature of the problem is not that different from some of those
constructions whose supersymmetry-breaking scale is at or above the
compactification scale\footnote{We are cautious in using the
  qualification ``some'' here. If the scale of supersymmetry is above
  the string scale, one would expect in addition to the supergravity
  fields that a tower of string states to come into play.}.  Studying
the effects of such supersymmetry-breaking backgrounds on the gauge
sector, irrespective of the origin of such breaking, will be one goal
of the present work

A particularly well-explored corner of $\cN_{4}=1$ constructions are
the class of flux compactifications of type-IIB string
theory~\cite{Verlinde:1999fy,*Dasgupta:1999ss,*Greene:2000gh,*Becker:1996gj,
  *Becker:2000rz,Giddings:2001yu} commonly known as GKP
compactifications\footnote{\label{foot:GKP}Strictly speaking, the
  compactifications of~\cite{Giddings:2001yu} are not necessarily
  supersymmetric.  However, since we are primarily interested in
  supersymmetric GKP compactifications, we will use the term ``GKP''
  to indicate the $\cN_{4}=1$ setups of~\cite{Giddings:2001yu}.
  Furthermore, GKP compactifications alone does not provide a
  mechanism for compactification, but our analysis will not depend on
  whether or not the K\"ahler structure moduli are stabilized.}.  This
class of constructions invokes closed-string flux which, in addition
to stabilizing many moduli, allows for constructing strongly warped
regions. Such strongly warped geometries provide a mechanism to
generate a hierarchy of scales via gravitational redshift, realizing
the bottom-up idea of~\cite{Randall:1999ee}.  This fact was exploited
in the KKLT construction~\cite{Kachru:2003aw}.  By combining this
strong warping with the quantum effects required to stabilize the
K\"ahler structure of the internal space, it was argued
in~\cite{Kachru:2003aw} that supersymmetry can be broken at a
hierarchically suppressed scale by the addition of a small number of
anti-branes which naturally inhabit points of strongest warping.  The
KKLT framework has been widely explored in the context of string
inflation (for reviews
see~\cite{Linde:2005dd,*Cline:2006hu,*Kallosh:2007ig,*Burgess:2007pz,
  *McAllister:2007bg,*Baumann:2009ni}) and in phenomenological
scenarios such as mirage mediation~\cite{Choi:2004sx,*Choi:2005ge} and
variations thereof~\cite{Everett:2008qy,*Everett:2008ey}.
Furthermore, the gauge/gravity correspondence is often realized with
strongly warped
geometries~\cite{Maldacena:1997re,*Gubser:1998bc,*Witten:1998qj,*Aharony:1999ti}.
The addition of a relatively small number of anti-branes to an
otherwise supersymmetric construction can, under certain
circumstances, be described as a meta-stable non-supersymmetric state
in a dual supersymmetric gauge
theory~\cite{Kachru:2002gs,DeWolfe:2008zy} .  This has been used to
construct gravity duals of gauge mediation
scenarios~\cite{Benini:2009ff,McGuirk:2009am,*Fischler:2011xd,*McGuirk:2011yg,*Skenderis:2012bs,*Argurio:2012cd}
(see also~\cite{Gabella:2007cp,*McGarrie:2010yk,*Okada:2011ed} for
related ideas) which are otherwise difficult to analyze using
conventional (perturbative) field theoretical techniques.  Finally,
anti-branes also play an important role in the large-volume
scenario~\cite{Balasubramanian:2005zx}, where non-perturbative effects
are played against $\ell_{\us}$-corrections to produce
intermediate-scale supersymmetry breaking without relying on strong
warping.

Despite the wide applications of this framework, the nature of
supersymmetry breaking by $\overline{\uD 3}$-branes remains somewhat
mysterious, even setting aside the subtleties involved in the
backreaction of the $\overline{\uD
  3}$s~\cite{DeWolfe:2008zy,McGuirk:2009xx,Bena:2009xk,*Bena:2011hz,*Bena:2011wh,*Massai:2012jn,Dymarsky:2011pm}.
In particular, it is not clear whether or not the breaking should be
considered explicit breaking or spontaneous breaking from the 4d point
of view, although the common folklore holds that it is the former.  An
argument often given for $\overline{\uD 3}$ branes providing an
explicit source of breaking is that the $\overline{\uD 3}$s preserve
the ``wrong'' supersymmetry, meaning the supersymmetry that is broken
by the D3 charge carried by the fluxes in a GKP compactification.
Indeed, such explicit breaking seems to be reflected in the effective
potential used in~\cite{Kachru:2003aw} in which the so-called uplift
potential, corresponding to the tension of the $\overline{\uD 3}$s, is
not included with the $F$-term potential\footnote{This of course
  leaves open the possibility that the $\overline{\uD 3}$ is a source
  of $D$-term breaking as suggested in, for
  example,~\cite{Camara:2003ku}.  We will provide some evidence for
  this possibility as well.}.

\newpage

On the other hand, the $\overline{\uD 3}$s can be thought of as a
soliton of closed strings, especially when the number of anti-branes
is large, in which case it is simply a non-supersymmetric
configuration in a supersymmetric theory; such a state of affairs is,
by definition, spontaneous breaking of supersymmetry.  In this sense,
the case of an anti-brane is quite similar to the case of branes
intersecting at angles.  Although for special angles, two intersecting
branes preserve some of the same supercharges, for generic angles they
will not.  One might be tempted to call this explicit breaking for
precisely the same reason as in the $\overline{\uD 3}$ case: at
generic angles the branes do not preserve the same supersymmetry.  Yet
since such angles are controlled by geometric and brane moduli, the
breaking by a non-trivial angle can be controlled by 4d fields and
therefore seems to be manifestly spontaneous breaking (see,
e.g.~\cite{Villadoro:2006ia} for related discussions). Indeed, this
was considered in, for example,~\cite{Antoniadis:2004uk} where the
theory for the corresponding goldstino, which indicates the
spontaneous breaking of supersymmetry, was discussed.  Since
$\overline{\uD p}$-branes differ from $\uD p$-branes precisely in
their orientation, the case of an $\uD p$-$\overline{\uD p}$ pair is
in some sense an extreme version of branes intersecting at
angles\footnote{This is admittedly a bit of a cheat; for example, for
  spacetime filling 3-branes transverse to a compact space, there is
  no finite-energy way to rotate the branes.}.  Finally, the case of
$\overline{\uD 3}$s in a GKP compactification is not intrinsically
distinct from the case of $\uD p$-branes in flat space as both involve
supercharges of 10d background being projected out by the localized
sources. In the latter case the massless scalars on the worldvolume
are the goldstones associated with the spontaneous breaking of
translational symmetry.  The massless fermions should be viewed in the
same light, as resulting from the spontaneous breaking of maximal
supersymmetry.  Indeed, the supersymmetric generalization of the
Dirac-Born-Infeld (DBI) action contains in it a Akulov-Volkov-like
(AV) action for goldstini~\cite{Volkov:1972jx,*Volkov:1973ix}.
Although this fact was understood long ago (see
e.g.~\cite{Aganagic:1996nn}) it seems, in our opinion, to be
under-appreciated.

In this work, we explore this question of explicit and spontaneous
breaking by considering non-supersymmetric perturbations to
supersymmetric GKP compactifications.  Although much of our analysis
is agnostic with respect to the source of these non-supersymmetric
perturbations, we have in mind those resulting from the addition of
$p$ $\overline{\uD 3}$s such that $p$ is much less than the number of
flux quanta that builds a warped region.  In generic cases, even
though only a single combination of fields is ``directly'' sourced by
the $\overline{\uD 3}$s, all other closed string fields are perturbed,
including non-Hermitian components of the internal metric.  As a
diagnostic of such breaking, we probe the resulting background with a
stack of $\uD 3$ branes and consider the resulting effective field
theory.  Such a situation has been considered previously in the
literature from both the
D-brane~\cite{Camara:2003ku,Camara:2004jj,Grana:2003ek,Burgess:2006mn}
and worldsheet points of view~\cite{Lust:2004fi,*Lust:2004dn}, but
none to our knowledge takes fully into account the non-Hermitian
perturbations to the internal metric (though see~\cite{Benini:2009ff}
for a related case) or explicitly analyze Yukawa couplings.  Although
this may seem like a slight distinction, the internal metric is the
matter-field metric for position moduli of the $\uD 3$ and so it
modifies the marginal operators (as well as operators of other
dimensions) of the $\uD 3$-brane effective field theory.  Since the
soft terms that result from the spontaneous breaking of supersymmetry
are all relevant operators (at least in the $m_{\mathrm{p}}\to\infty$
limit), this would seem to hint at explicit breaking.  Nevertheless,
we find that a simple non-holomorphic field redefinition puts the
effective field theory into a form that manifestly exhibits only soft
breaking. As generic explicit breaking should lead to hard terms, even
in the limit as $m_{\mathrm{p}}\to\infty$, we take this as an
indication that the breaking of supersymmetry may be spontaneous.

Let us stress that since $\uD 3$s are local objects, the analysis of
the $\uD 3$ action is, through marginal order, fairly insensitive to
the form that the internal metric takes (so long as it is not
singular) and previous analyses of the $\uD 3$-action are
straightforwardly adopted to the case of a general metric.  Indeed
though the analyses of~\cite{Camara:2003ku, Grana:2003ek} take the
ansatz where the internal metric remains Calabi-Yau, their results are
largely valid in more general cases\footnote{Notable exceptions are
  the non-renormalizable couplings between open and closed strings
  considered in~\cite{Grana:2003ek} which depend on an understanding
  of the light closed-string spectrum that is not available in
  general.} except for the fact that they rely on the underlying
Calabi-Yau to give a complex structure to the open-string effective
field theory.  In this light, our goal is to not to greatly extend the
technical advances of these works, but instead to make steps towards a
conceptual understanding of supersymmetry breaking.

We also emphasize that even though we primarily work in the context of a
non-supersymmetric perturbation to GKP, the $\uD 3$-brane Lagrangian
seems to be soft independently of the background or even if the
closed-string equations of motion are applied.  However, in the case
in which the background is a result of the backreaction of
$\overline{\uD 3}$s in GKP, we are also able to identify the gaugino living on
the $\overline{\uD 3}$-brane as a candidate for the goldstino that is
expected to be present if supersymmetry is spontaneously broken.  For
this reason, much of our discussion is framed within the context of
supersymmetry breaking by the addition of anti-branes.

This paper is organized as follows.  In section~\ref{sec:GKP}, we
review GKP compactifications and argue that the addition of an
$\overline{\uD 3}$ brane will generically perturb all closed string
fields including the internal metric.  In section~\ref{sec:D3_EFT}, we
discuss the effective field theory of a stack of $\uD 3$-branes or
$\overline{\uD 3}$-branes probing such a geometry.  In
section~\ref{sec:soft}, we review the nature of soft breaking of
supersymmetry and show how the action presented in the previous
section falls into this class, though the supersymmetry that is
``least'' broken is not quite that preserved by GKP.  In
section~\ref{sec:goldstino}, we discuss a candidate for a goldstino
field on an $\overline{\uD 3}$ probing a GKP compactification.  Some
concluding remarks are given in section~\ref{sec:conclusion} and our
conventions are summarized in appendix~\ref{app:conv}.

\section{\label{sec:GKP}Non-supersymmetric perturbations to GKP compactifications}

In this section, we discuss non-supersymmetric perturbations to
$\cN_{4}=1$ GKP compactifications~\cite{Giddings:2001yu} of the
type-IIB superstring, with an emphasis on those resulting from the
addition of a number of $\overline{\uD 3}$-branes that is small
compared to the amount of flux in the supersymmetric case.  GKP
compactifications are of the form $R^{3,1}\times_{\mathrm{w}} X^{6}$
where $\times_{\mathrm{w}}$ indicates a non-trivial fibration of
$R^{3,1}$ over the compact internal space $X^{6}$.  The metric takes
the familiar warped ansatz
\begin{subequations}
\label{eq:GKP_ansatz}
\begin{equation}
  \ud s_{10}^{2}=\hat{g}_{MN}\ud x^{M}\ud x^{N}
  =\ue^{2A\left(y\right)}\eta_{\mu\nu}\ud x^{\mu}\ud x^{\nu}+
  \ue^{-2A\left(y\right)}{g}_{mn}\ud y^{m}\ud y^{n}.
\end{equation}
The geometry is supported by a 3-form flux
$G_{\left(3\right)}=F_{\left(3\right)}+\ui\ue^{-\phi}H_{\left(3\right)}$
without legs on the external space $R^{3,1}$ and a 5-form flux
\begin{equation}
  F_{\left(5\right)}=\bigl(1+\hat{\ast}\bigr)\cF_{\left(5\right)},\qquad
  \cF_{\left(5\right)}=\ud\alpha\wedge\ud\vol_{R^{3,1}},
\end{equation}
\end{subequations}
in which $\ud\vol_{R^{3,1}}$ is the volume form for $R^{3,1}$ and
$\hat{\ast}$ is the 10d Hodge-$\ast$ built from the metric
$\hat{g}_{MN}$.  Our interest is in the regime where dimensional
reduction on the $X^{6}$ produces an effective 4d theory.  Such a
theory will exhibit $\cN_{4}\ge 1$
if~\cite{Grana:2001xn,Giddings:2001yu}
\begin{enumerate}
\item $X^{6}$ is a K\"ahler manifold and $g_{mn}$ is the associated
  K\"ahler metric,
\item the 3-form flux is primitive and has Hodge type
  $\left(2,1\right)$ and is therefore imaginary self-dual (ISD), $\ui
  G_{\left(3\right)}=\ast G_{\left(3\right)}$, where $\ast$ (without
  the hat) denotes the 6d Hodge-$\ast$ built from ${g}_{mn}$,
\item the $5$-form flux and the warp factor are related by
  $\ue^{4A}=\alpha$,
\item the axiodilaton $\tau=C_{\left(0\right)}+\ui\ue^{-\phi}$ varies
  holomorphically over $X^{6}$.
\end{enumerate}
A construction satisfying these requirements is called a GKP
compactification (though see footnote~\ref{foot:GKP}).  These
compactifications must in addition contain certain sources (D3-branes,
O3-branes, or 7-branes) to ensure the cancellation of tadpoles;
however, these sources will not play a significant role in our
analysis.

As reviewed in the introduction, GKP compactifications are a
particularly interesting region of the landscape since, while they are
based upon the comparatively well-understood K\"ahler and Calabi-Yau
geometries, the presence of non-trivial 3-form flux can stabilize the
complex structure of $X^{6}$, the axiodilaton, and the deformation
moduli for $7$-branes. Additionally, these constructions can
accommodate low-scale supersymmetry breaking as large amounts of flux
can produce strongly warped regions.  Since ISD flux carries ${\uD 3}$
charge, $\overline{\uD 3}$s, which carry the opposite-sign charge and
hence break the supersymmetry preserved by GKP, are naturally
attracted to the regions of strongest warping and so the corresponding
scale of supersymmetry breaking can be highly redshifted.  In some
cases, when $\uD 3$-branes are absent, the $\overline{\uD 3}$s will be
perturbatively stable, only decaying into flux and $\uD 3$-branes
after undergoing a Myers-like effect~\cite{Myers:1999ps} followed by a
quantum tunneling process~\cite{Kachru:2002gs}.

In order to perform a detailed study of such constructions, the
influence of such $\overline{\uD 3}$-branes on the background must be
considered.  The most studied example is the Klebanov-Strassler (KS)
geometry~\cite{Klebanov:2000hb} which results from ISD 3-form flux
threading the deformed conifold.  The backreaction of a small number
of $\overline{\uD 3}$s on the KS geometry has been a topic of recent
interest~\cite{DeWolfe:2008zy,McGuirk:2009xx,Bena:2009xk,Dymarsky:2011pm,*Bena:2011hz,*Massai:2012jn}.
Due to the presence of the background 3-form flux of KS, the addition
of the $\overline{\uD 3}$s produces a non-ISD flux and in fact, near
the anti-branes, all Hodge types of $3$-form flux are
present~\cite{McGuirk:2009xx}. Furthermore, it was pointed out
in~\cite{Benini:2009ff} that the $\overline{\uD 3}$s perturb the
metric in such a way that the internal metric $g_{mn}$ is no longer
Hermitian with respect to the original complex structure: the
backreaction of the $\overline{\uD 3}$s includes non-vanishing metric
components $g_{zz}$ and $g_{\bar{z}\bar{z}}$ when expressed in terms
of the complex coordinates of the original deformed conifold.

The fact that such non-Hermitian components will generically appear
after the addition of $\overline{\uD 3}$s can be easily seen from the
type-IIB equations of motion.  Let us again consider the
ansatz~\eqref{eq:GKP_ansatz} but relax the conditions for
supersymmetry.  It is useful to construct the combinations
\begin{equation}
  \Phi_{\pm}=\ue^{4A}\pm\alpha,\quad
  G_{\pm}=\bigl(\ast_{6}\pm \ui\bigr)G_{\left(3\right)},\quad
  \Lambda=\Phi_{+}G_{-}+\Phi_{-}G_{+}.
\end{equation}
The equations of motion and Bianchi identities~\eqref{eq:IIB_eom} can
be expressed in these fields as~\cite{Giddings:2001yu,Baumann:2010sx}
\begin{subequations}
\begin{align}
  0=&\nabla^{2}\Phi_{\pm}
  -\frac{\bigl(\Phi_{+}+\Phi_{-}\bigr)^{2}}{16\,\mathrm{Im}\,\tau}
  \left\lvert G_{\pm}\right\rvert^{2}
  -\frac{2}{\Phi_{+}+\Phi_{-}}\left\lvert\partial\Phi_{\pm}\right\rvert^{2},\\
  0=&\ud\Lambda+\frac{\ui}{2\,\mathrm{Im}\,\tau}
  \ud\tau\wedge\bigl(\Lambda+\xoverline{\Lambda}\bigr),\\
  0=&\ud\bigl(G_{\left(3\right)}-\tau H_{\left(3\right)}\bigr),\\
  0=&\nabla^{2}\tau+\frac{\ui}{
    \mathrm{Im}\,\tau}{\left(\partial\tau\right)^{2}}
  +\frac{\ui}{8}\bigl(\Phi_{+}+\Phi_{-}\bigr)G_{+}\cdot G_{-},\\
  0=&R_{mn}-\frac{1}{2\left(\mathrm{Im}\,\tau\right)^{2}}
  \partial_{\left(m\right.}\tau\partial_{\left.n\right)}\bar{\tau}
  -\frac{2}{\left(\Phi_{+}+\Phi_{-}\right)^{2}}
  \partial_{\left(m\right.}\Phi_{+}\partial_{\left.n\right)}\Phi_{-}\notag\\
  &+\frac{\Phi_{+}+\Phi_{-}}{16\cdot 2!\, \mathrm{Im}\,\tau}
  \biggl[G_{+\left(m\right.}^{\phantom{+\left(m\right.}pq}\,
  \xoverline{G}^{\phantom{\left(+p\right.}}_{-\left.n\right)pq}
  +G_{-\left(m\right.}^{\phantom{+\left(m\right.}pq}\,
  \xoverline{G}^{\phantom{+\left(p\right.}}_{+\left.n\right)pq}\biggr],
\end{align}
\end{subequations}
in which, for simplicity of presentation, we have omitted terms
resulting from localized sources.  For $p$-forms we use the notation
\begin{equation}
  \label{eq:define_dot_product}
  X_{\left(p\right)}\cdot Y_{\left(p\right)}
  =\frac{1}{p!}X_{m_{1}\cdots m_{p}}Y^{m_{1}\cdots m_{p}},\qquad
  \left\lvert X_{\left(p\right)}\right\rvert^{2}=X_{\left(p\right)}
  \cdot\overline{X}_{\left(p\right)}.
\end{equation}
Note that we have defined
$\overline{G}_{\pm}=\bigl(G_{\pm}\bigr)^{\ast}$ so that, for example,
$\xoverline{G}_{+}$ is imaginary anti-self-dual (IASD). Here and
throughout we perform contractions and construct connections with the
unwarped metric $g_{mn}$ unless otherwise noted.  $\cN_{4}\ge 0$ GKP
compactifications are characterized by the conditions $\Phi_{-}=0$ and
$G_{-}=0$ with $\cN_{4}\ge 1$ having the additional requirement that
$G_{\left(3\right)}$ is a primitive $\left(2,1\right)$-form.  For
non-vanishing $\Phi_{+}$, we can recast the equation of motion for
$\Phi_{+}$ as~\cite{Gandhi:2011id}
\begin{equation}
  0=\nabla^{2}\Phi_{+}^{-1}
  +\frac{1}{16\,\mathrm{Im}\,\tau}
  \frac{\left(\Phi_{+}+\Phi_{-}\right)^{2}}{\Phi_{+}^{2}}
  \left\lvert G_{+}\right\rvert^{2}
  +\frac{2}{\Phi_{+}}\biggl[\frac{1}{\left(\Phi_{+}+\Phi_{-}\right)}
  -\frac{1}{\Phi_{+}}\biggr]
  \bigl(\partial\Phi_{+}\bigr)^{2}.
\end{equation}
    
For the moment, we will specialize to the case in which we start with
$G_{-}=0$, $\Phi_{-}=0$ and $\tau$ is a constant so that $X^{6}$ is a
Calabi-Yau.  We can then consider a perturbation such as, for example,
the addition of a small number of $\overline{\uD 3}$-branes.  Then
remarkably the linearized equations of motion for the perturbations
take a nearly triangular form~\cite{Gandhi:2011id}
\begin{subequations}
\label{eq:toolkit}
\begin{align}
  \nabla^{2}\delta\Phi_{-}=&0,\\
  \ud\bigl(\Phi_{+}\delta G_{-}\bigr)=&-\ud\bigl(\delta\Phi_{-}G_{+}\bigr),\\
  \bigl(\ast+\ui\bigr)\delta G_{-}=&0,\\
  \nabla^{2}\delta\tau=&-\frac{\ui}{8}\Phi_{+}\bigl(G_{+}\cdot\delta G_{-}\bigr),
  \\
  \label{eq:toolkit_metric}
  -\frac{1}{2}\Delta \delta g_{mn}=&\frac{2}{\Phi_{+}^{2}}
  \partial_{\left(m\right.}\Phi_{+}\partial_{\left. n\right)}\delta\Phi_{-}
  -\frac{\Phi_{+}}{16\cdot 2!}
  \biggl[G_{+\left(m\right.}^{\phantom{+\left(m\right.}pq}\,
  \delta\xoverline{G}^{\phantom{p}}_{-\left.n\right)pq}
  +\delta G_{-\left(m\right.}^{\phantom{+\left(m\right.}pq}\,
  \xoverline{G}^{\phantom{p}}_{+\left.n\right)pq}\biggr],\\
  \ud\delta G_{+}=&\ud\bigl(\delta G_{-}+2\ui\delta\tau H_{\left(3\right)}\bigr),\\
  \bigl(\ast-\ui\bigr)\delta G_{+}=&0,\\
  -\nabla^{2}\delta \Phi_{+}^{-1}=&
  \bigl(\delta\nabla\bigr)^{2}\Phi_{+}^{-1}
  -\frac{1}{16}\mathrm{Im}\,\delta\tau
  \left\lvert G_{+}\right\vert^{2}\notag\\
  &+\frac{1}{16}\biggl[G_{+}\cdot\delta\xoverline{G}_{+}
  +\delta G_{+}\cdot \xoverline{G}_{+}
  +\frac{1}{2!}G_{+\, m_{1}n_{1}p_{1}}\xoverline{G}_{+\, m_{2}n_{2}p_{2}}
  g^{m_{1}m_{2}}g^{n_{1}n_{2}}\delta g^{p_{1}p_{2}}\biggr]\notag\\
  &+\biggl[\frac{1}{8}\Phi_{+}^{-1}\left\lvert G_{+}\right\rvert^{2}
  -2\Phi_{+}^{-4}\bigl(\partial \Phi_{+}\bigr)^{2}\biggr]
  \delta\Phi_{-}.
\end{align}
\end{subequations}
Here $\delta\Psi$ denotes a perturbation to a field,
$\Psi\to\Psi+\delta \Psi$ and
\begin{equation}
  \Delta\delta g_{mn}:=
  \nabla^{2}\delta g_{mn}
  +\nabla_{m}\nabla_{n}\bigl(g^{pq}\delta g_{pq}\bigr)
  -2\nabla^{p}\nabla_{\left(m\right.}\delta g_{\left.n\right)p}.
\end{equation}
We have additionally set the unperturbed constant axiodilaton to
$\tau=\ui$ and again omitted the explicit appearances of source terms.
Although we will not make use of it, this pattern of triangularity
continues order-by-order in perturbation theory\footnote{We note that
  the equations of motion may not always be truly triangular.  For
  example, in general the metric is characterized by many functions
  and~\eqref{eq:toolkit_metric} will generically not have any special
  structure for those functions.  This is the case for perturbations
  to the KS geometry~\cite{Bena:2009xk} except in the nearly-conformal
  region~\cite{Gandhi:2011id}.}.

This form of the equations of motion is useful since it is precisely
the mode $\Phi_{-}$ that is ``directly'' sourced by a $\overline{\uD
  3}$-brane in the sense that only the equation of motion for
$\Phi_{-}$ has a $\delta$-function term in the presence of
$\overline{\uD 3}$s.  That an $\overline{\uD 3}$ sources $\Phi_{-}$
can be most easily seen by placing an $\overline{\uD 3}$ in flat space
where the only field that becomes non-trivial is $\Phi_{-}$.
From~\eqref{eq:toolkit}, we see that in the presence of $G_{+}\neq 0$,
once $\delta \Phi_{-}$ is non-zero, $\delta G_{-}$ is non-zero as
well, and indeed $\delta G_{-}$ generically possess all Hodge
types\footnote{This genericity is violated in, for
  example,~\cite{DeWolfe:2008zy} where the imposed R-symmetry requires
  $G_{\left(3,0\right)}=0$ (where the Hodge-type is given in terms of
  the original complex structure).}. The presence of $\delta G_{-}\neq
0$ gives a source for $\delta\tau$ and generically both the real and
imaginary components are non-vanishing\footnote{For the example of KS,
  $\tau$ is pure imaginary after the addition of the $\overline{\uD
    3}$ brane since $H_{\left(3\right)}$ and $F_{\left(3\right)}$
  thread dual cycles and so $F_{\left(3\right)}\cdot
  H_{\left(3\right)}\propto \mathrm{Im}\left(G_{+}\cdot
    G_{-}\right)=0$ automatically, even after the perturbation.}.
Inserting the directly sourced $\delta\Phi_{-}$ and the indirectly
sourced $\delta G_{-}$ into the equation for $\delta g_{mn}$
generically forces all components to be non-vanishing.  For example, a
$\left(2,1\right)$ $G_{+}$ and $\left(3,0\right)$ $\delta G_{-}$ act
as a source $\delta g_{\bar{z}\bar{z}}$
component\footnote{In~\cite{DeWolfe:2008zy}, there was a non-vanishing
  $\delta g_{\bar{z}\bar{z}}$ even though no $\left(3,0\right)$ flux
  was sourced.  This is because the left-hand side
  of~\eqref{eq:toolkit_metric} involves all components of the
  perturbed metric and so even sourcing the $\bar{z}z$ component of
  $\Delta\delta g_{mn}$ will generically result in non-vanishing
  $\delta g_{\bar{z}\bar{z}}$.}.  Similarly, $\Phi_{+}$ and $\Phi_{-}$ are real functions
and so $\delta\Phi_{-}\neq 0$ should also generically source all
components of the metric\footnote{Note that at least in some simple
  fluxless cases such as a $\uD 3$-$\overline{\uD 3}$ pair in flat
  space, we can choose a coordinate system such that the internal
  space is still Hermitian with respect to the original complex
  structure, but at the expense of having a different scaling factor
  for the transverse
  metric~\cite{DeWolfe:2008zy,Zhou:1999nm,*Brax:2000cf}.}.  Following
the remainder of the equations as above also leads us to conclude that
$G_{+}$ and $\Phi_{+}$ are perturbed from the original background
values.  We note also that this argument implies that an initial
singularity in $\delta \Phi_{-}$, such as that appearing
in~\cite{DeWolfe:2008zy,McGuirk:2009xx,Bena:2009xk}, is felt by all
perturbed fields, even if $\Phi_{-}$ is the only field directly
sourced.

The presence of the singularities in the fields not directly sourced
by the $\overline{\uD 3}$s is perhaps surprising and has been a topic
of recent
discussion~\cite{Bena:2009xk,Bena:2011hz,*Bena:2011wh,*Massai:2012jn,Blaback:2011nz,*Blaback:2011pn,*Blaback:2012nf,
  *Bena:2012}.  The $\overline{\uD 3}$s directly source $\delta
\Phi_{-}$ and so the corresponding divergence is as physically
acceptable as the divergence in the electric field at the position of
a point-charge in classical Maxwell theory.  In contrast, the $3$-form
flux and other fields are not directly sourced by these fields and so
the corresponding singularities might be seen as suspect.  Here we
take the point of view that, due to the non-linearity of the
supergravity equations of motion and the fact that all of the fields
couple to each other, once one sort of singularity is accepted,
divergences in all other fields must be accepted as well.  Indeed,
presumably there exists some stringy mechanism that resolves the
singularity in $\Phi_{-}$ (for example, an $\overline{\uD 3}$, even in
flat space, should have some finite width comparable to the string
length) and once $\Phi_{-}$ is rendered finite, there is no reason to
expect that any of the other singularities will be present (however,
the linearized analysis of the supergravity equations of motion is
expected to be inapplicable).  We therefore view it is as plausible
that the divergences will be resolved in a full treatment and so
accept the apparent singularities as being a consequence of an
incomplete treatment (see also~\cite{Dymarsky:2011pm} for responses to
the objections related to these divergences).  Nevertheless, because
supergravity may break down near the position of $\overline{\uD 3}$s,
we will assume in what follows that we are evaluating our fields
sufficiently far away from any such sources.

To summarize, we have argued that in a generic $\cN_{4}=1$ GKP
compactification, the addition of an $\overline{\uD 3}$-brane will
cause the configuration to move away from all of the supersymmetry
conditions, perturbing all Hodge-types of flux, causing the
axiodilaton to be non-vanishing, and forcing the internal metric to be
no-longer Hermitian, even though the $\overline{\uD 3}$ itself
directly sources only $\Phi_{-}$.  For simplicity and since the
equations of motion are almost triangular, we have worked in the
special case in which the axiodilaton is constant in the unperturbed
geometry.  However, it would be rather surprising if in the more
generic case of varying axiodilaton that these perturbations were not
produced.  Hence, it what follows we will drop the assumption of
constant axiodilaton.  Further, although we have emphasized in this
section perturbations due to the presence of $\overline{\uD
  3}$-branes, the analysis of the $\uD 3$-action will be independent
of the source of these perturbations, though we will assume that
supergravity is still applicable.

Note that in the above discussion, we have neglected the influence of
the non-perturbative effects that are required to stabilize the
K\"ahler structure~\cite{Kachru:2003aw}.  Such non-perturbative
reactions will backreact on the geometry in such a way that it will be
better described as a generalized complex
geometry~\cite{Koerber:2007xk,Heidenreich:2010ad,Baumann:2010sx}.
Although such effects might naively seem to be negligible, they may
spoil important properties such as sequestering~\cite{Berg:2010ha}.
In principle, we could try to fold the backreaction of the
non-perturbative effects into the perturbations of GKP that in the
above we attributed to the supersymmetry-breaking sources.  However,
the points that we wish to make are independent of whether or not the
K\"ahler structure is in fact stabilized and so we will leave the
incorporation of such effects for future work.

\section{\label{sec:D3_EFT}Effective action for D3s}

In this section, we consider the effective action for a stack of
coincident $\uD 3$-branes probing a perturbation to an $\cN_{4}\ge1$
GKP compactification. Our analysis is similar to that performed
in~\cite{Camara:2003ku,Grana:2003ek,Burgess:2006mn} (see
also~\cite{Lust:2004fi,*Lust:2004dn}) and indeed we recover many of
the same results, except that we take into account the fact that
non-supersymmetric fluxes will generically cause the internal metric
to no longer be Hermitian with respect to the unperturbed complex
structure.  We perform the analysis for both probe $\uD 3$-branes and
$\overline{\uD 3}$-branes but will frequently, in this section, use
``$\uD 3$'' to denote a 3-brane of either charge.  In
section~\ref{sec:soft}, we will re-express the resulting action for a
$\uD 3$ in the language of softly-broken supersymmetry and comment on
how our results relate to those appearing elsewhere in the literature.

\subsection{\label{sec:bosonic}Bosonic action}

The effective action for the light open-string bosonic fluctuations of
a single $\uD p$-brane in either type-II string theory consists of the
familiar DBI and Chern-Simons (CS) terms which in
the 10d Einstein frame take the form
\begin{subequations}
\begin{align}
  S_{\uD p}=&S_{\uD p}^{\mathrm{DBI}}+S_{\uD p}^{\mathrm{CS}},\\
  S_{\uD p}^{\mathrm{DBI}}=&
  -\tau_{\uD p}\int\ud^{p+1}\xi\,\ue^{\frac{p-3}{4}\phi}
  \sqrt{-\det\bigl(\hat{M}_{\alpha\beta}\bigr) },\\
  S_{\uD p}^{\mathrm{CS}}=&
  \pm\tau_{\uD p}\int \mathrm{P}\biggl[\sum_{n}C_{\left(n\right)}
  \wedge
  \ue^{B_{\left(2\right)}}\biggr]\wedge \ue^{\ell_{\us}^{2}f_{\left(2\right)}},
\end{align}
\end{subequations}
where the upper (lower) sign applies for a $\uD p$-brane
($\overline{\uD p}$-brane).  The integral is over the worldvolume of
the brane and the tension and charge of a $\uD p$ brane are given by
$\tau_{\uD p}^{-1}=\frac{1}{2\pi}\ell_{\us}^{p+1}g_{\us}$.  Away from
orientifold planes, the bosonic fields consist of a $\U{1}$
gauge-field $A_{\left(1\right)}$ with field strength
$f_{\left(2\right)}=\ud A_{\left(1\right)}$ and the transverse
deformations which enter through the pullback of bulk fields to the
worldvolume denoted by $\mathrm{P}$.  $\xi^{\alpha}$ are the
worldvolume coordinates and choosing the static gauge we have
\begin{equation}
  \mathrm{P}\bigl[v_{\alpha}\bigr]=v_{\alpha}
  +\ell_{\us}^{2}v_{i}\partial_{\alpha}\vp^{i},
\end{equation}
where we have defined the worldvolume scalars
$\vp^{i}=\ell_{\us}^{-2}X^{i}$ in which $X^{i}$ are coordinates
transverse to the worldvolume.  Here we have defined
\begin{equation}
\label{eq:generalized_metric}
  \hat{M}_{\alpha\beta}=\mathrm{P}\bigl[\hat{E}_{\alpha\beta}\bigr]
  +\ue^{-\phi/2}\ell_{\us}^{2}f_{\alpha\beta},\qquad
  \hat{E}_{MN}=\hat{g}_{MN}+\ue^{-\phi/2}B_{MN}.
\end{equation}

For a stack of $N$ $\uD p$-branes the gauge symmetry on the common
worldvolume is promoted to a $\U{N}$ gauge symmetry and the transverse
deformations $\vp^{i}$ are promoted to adjoint-valued fields.  The DBI
and CS actions then become modified to~\cite{Myers:1999ps}
\begin{subequations}
\label{eq:Myers}
\begin{align}
\label{DBI}
  S_{\uD p}^{\mathrm{DBI}}=& 
  -\tau_{\uD p}
  \int\ud^{p+1}\xi\,
  \Str\biggl\{\ue^{\frac{p-3}{4}\phi}
  \sqrt{-\det\bigl(\hat{M}_{\alpha\beta}\bigr)
  \det\bigl(Q^{i}_{\, j}\bigr)}\biggr\},\\
\label{CS}
  S_{\uD p}^{\mathrm{CS}}=&
  \pm\tau_{\uD p}\int\Str\biggl\{
  \mathrm{P}\biggl[\ue^{\ui\ell_{\us}^{2} \iota^{2}_{\varphi}}\biggl(
  \sum_{n}C_{\left(n\right)}\wedge \ue^{B_{\left(2\right)}}\biggr)
  \biggr]\wedge \ue^{\ell_{\us}^{2} f_{\left(2\right)}}\biggr\}.
\end{align}
\end{subequations}
In the static gauge in which we work, we redefine
\begin{equation}
\label{M}
  \hat{M}_{\alpha\beta}=
  \mathrm{P}\left[\hat{E}_{\alpha\beta}+\ue^{\phi/2}\hat{E}_{\alpha i}
    \left(Q^{-1}-\delta\right)^{ij}\hat{E}_{j\beta}\right]
  + \ue^{-\phi/2}\ell_{\us}^{2}f_{\alpha\beta},
\end{equation}
in which
\begin{equation}
  Q^{i}_{\, j}=\delta^{i}_{\, j}
  +\ui\ell_{\us}^{2}
  \bigl[\varphi^{i},\varphi^{k}\bigr]\ue^{\phi/2}\hat{E}_{kj}.
\end{equation}
The field strength is modified to $f_{\left(2\right)}=\ud
A_{\left(1\right)}-\ui A_{\left(1\right)}\wedge A_{\left(1\right)}$
and the pullback to a non-Abelian pullback
\begin{equation}
  \mathrm{P}\bigl[v_{\alpha}\bigr]=v_{\alpha}
  +\ell_{\us}^{2}v_{i}D_{\alpha}\vp^{i},
\end{equation}
where
\begin{equation}
  D_{\alpha}=\partial_{\alpha}-\ui\bigl[A_{\alpha},\cdot\bigr],
\end{equation}
is the usual gauge-covariant derivative acting on adjoint-valued
fields.  $\iota_{\vp}$ denotes an interior product,
\begin{equation}
  \iota_{\vp}\bigl(v_{M}\ud x^{M}\bigr)=\vp^{i}v_{i},\qquad
  \iota_{\vp}^{2}\bigl(\frac{1}{2}v_{MN}\ud x^{M}\ud x^{N}\bigr)
  =\frac{1}{2}\bigl[\vp^{j},\vp^{i}\bigr]v_{ij}.
\end{equation}
Note that due to the non-Abelian nature of the theory,
$\iota_{\vp}^{2}\neq 0$.  A bulk field appearing in the D-brane action
is to be interpreted as a non-Abelian Taylor expansion,
\begin{eqnarray}\label{Taylor exp}
  \Psi\bigl(\varphi\bigr)=\sum_{n=0}^{\infty}\frac{\ell_{\us}^{2n}}{n!}
  \varphi^{i_{1}}\cdots\varphi^{i_{n}}
  \biggl[\partial_{i_{1}}\cdots\partial_{i_{n}}\Psi\bigl(\varphi\bigr)
  \biggr]_{\varphi=0}.
\end{eqnarray}
Finally, $\mathrm{Str}$ denotes a particular trace
prescription~\cite{Myers:1999ps}: before tracing over gauge indices,
the expression is symmetrized over factors of $f_{\alpha\beta}$,
$D_{\alpha}\vp^{i}$, $\left[\vp^{i},\vp^{j}\right]$ and the $\vp^{i}$
appearing in the Taylor expansion.  Note that this allows us to treat
these objects as commuting.

Our goal is to deduce the effective action to leading order in
$\ell_{\us}$.  That is, the bosonic action consists of an infinite
series of irrelevant operators that can be thought of as arising from
integrating out massive string modes.  Since our interest is the
long-wavelength theory, we will consider only the relevant and
marginal operators.  Furthermore, the coefficients of these operators
generically have expansions of the schematic form
\begin{equation}
  \label{eq:coefficient_sum}
  c\sim\sum_{n}\ell_{\us}^{n}\partial^{n}\Psi,
\end{equation}
in which $\Psi$ indicates a bulk field and $\partial^{n}$ indicates
$n$ derivatives.  Since we wish to work in the supergravity regime, we
must consider backgrounds where $\ell_{\us}$ corrections to the
supergravity action~\eqref{eq:IIB_eom} can be neglected, and thus we
must have that the derivatives of bulk fields are small with respect
to the string scale, at least when evaluated near the position of the
probe branes.  Therefore we can truncate the
sum~\eqref{eq:coefficient_sum} after a certain number of terms.  Note
that for both expansions, we are comparing energies to
$\ell_{\us}^{-1}$ and so, although it's dimensionful, we can expand in
powers of $\ell_{\us}$ as a proxy for the double expansion in powers
of open-string fields and closed-string curvatures.  As evidenced by
explicit
examples~\cite{DeWolfe:2008zy,McGuirk:2009xx,Bena:2009xk,Bena:2011hz,*Bena:2011wh,*Massai:2012jn,Dymarsky:2011pm,Blaback:2011nz,*Blaback:2011pn,*Blaback:2012nf,
  *Bena:2012} and discussed in the previous section, the closed-string
background will generically be divergent at the position of the
$\overline{\uD 3}$s (at least in the approximation of linearized
supergravity) and so we will work far from the anti-branes.

For the case of interest $p=3$ and we can replace the worldvolume
indices $\alpha, \beta$ with the usual $R^{3,1}$ indices $\mu, \nu$ and
the transverse indices with the internal indices of $X^{6}$ $m, n$.  Then
\begin{equation}
  \hat{M}_{\mu\nu}=\ue^{2A\left(\vp\right)}\eta_{\mu\nu}
  +\ue^{-\phi\left(\vp\right)/2}\ell_{\us}^{2}f_{\mu\nu}
  +\ell_{\us}^{4}\bigl(\ue^{-2A\left(\vp\right)}g_{mn}\bigl(\vp\bigr)+
  \ue^{-\phi\left(\vp\right)/2}B_{mn}\bigl(\vp\bigr)\bigr)
  D_{\mu}\vp^{m}D_{\nu}\vp^{n},
\end{equation}
since $B_{\mu\nu}=0$, $\hat{E}_{\mu m}=0$ and
$Q=1+\cO\left(\ell_{\us}^{2}\right)$.  Then, making use of the
identity
\begin{equation}
  \sqrt{\det(1+M)}=
  1+\frac{1}{2}\tr\bigl(M\bigr)-\frac{1}{4}\tr\bigl(M^{2}\bigr)
  +\frac{1}{8}\bigl[\tr\bigl(M\bigr)\bigr]^{2}+\cdots,
\end{equation}
we have
\begin{equation}
  \sqrt{-\det\bigl(\hat{M}_{\mu \nu}\bigr)}=
  \ue^{4A\left(\vp\right)}+
  \frac{\ell_{\us}^{4}}{2}g_{mn}\left(\vp\right)D_{\mu}\vp^{m}D^{\mu}\vp^{n}
  +\frac{\ell_{\us}^{4}\ue^{-\phi\left(\vp\right)}}{4}f_{\mu\nu}f^{\mu\nu},
\end{equation}
where we have made use of the anti-symmetry of $f_{\left(2\right)}$
and $B_{\left(2\right)}$.  From this expression, we see that we are
interested in an expansion through $\cO\left(\ell_{\us}^{4}\right)$.
Performing the Taylor expansions,
\begin{align}
  \sqrt{-\det\bigl(\hat{M}_{\mu \nu}\bigr)}
  =&\frac{1}{2}\bigl(\Phi_{+}+\Phi_{-}\bigr)
  +\frac{\ell_{\us}^{4}}{2}g_{mn}D_{\mu}\vp^{m}D^{\mu}\vp^{n}
  +\frac{\ell_{\us}^{4}\, \mathrm{Im}\, \tau}{4}f_{\mu\nu}f^{\mu\nu}\notag\\\
  &+\frac{\ell_{\us}^{2}}{2}\partial_{m}\bigl(\Phi_{+}+\Phi_{-}\bigr)\vp^{m}
  +\frac{\ell_{\us}^{4}}{4}\partial_{m}\partial_{n}
  \bigl(\Phi_{+}+\Phi_{-}\bigr)\vp^{m}\vp^{n}.
\end{align}
Our notation is such that closed-string fields without an expressed
$\vp$ dependence are to be evaluated at $\vp=0$ and external indices
are contracted only with $\eta^{\mu\nu}$.  Note that the Taylor
expansion of the warp factor demonstrates the point discussed
regarding~\eqref{eq:coefficient_sum}: further expansion leads to
operators that are relevant and marginal (as well as irrelevant), but
their coefficients are suppressed by higher orders in derivatives of
the warp factor which we take to be small compared to the string
scale.

Similarly,
\begin{align}
  \sqrt{\det\bigl(Q^{m}_{\phantom{m}n}\bigr)}=&1-
  \frac{\ui \ell_{\us}^{2}}{2}B_{mn}\bigl(\vp\bigr)
  \bigl[\varphi^{m},\varphi^{n}\bigr]
  -\frac{\ell_{\us}^{4}}{4}\ue^{-4A\left(\vp\right)}\ue^{\phi\left(\vp\right)}
  g_{mn}\bigl(\vp\bigr)g_{pq}\bigl(\vp\bigr)
  \bigl[\vp^{m},\vp^{p}\bigr]\bigl[\vp^{n},\vp^{q}\bigr]\notag\\
  &-\frac{\ell_{\us}^{4}}{8}
  B_{mn}\bigl(\vp\bigr)B_{pq}\bigl(\vp\bigr)\bigl[\vp^{m},\vp^{n}\bigr]\bigl[\vp^{p},\vp^{q}\bigr],
\end{align}
where we have made use of the symmetry properties of $g_{mn}$ and
$B_{mn}$. We can choose a gauge so that $B_{\left(2\right)}=0$ at
$\vp=0$, and so this result simplifies to
\begin{equation}
  \sqrt{\det\bigl(Q^{m}_{\phantom{m}n}\bigr)}
  =1-\frac{\ui\ell_{\us}^{4}}{2}\partial_{p}B_{mn}\vp^{p}
  \bigl[\vp^{m},\vp^{n}\bigr]
  -\frac{\ell_{\us}^{4}}{2\left(\Phi_{+}+\Phi_{-}\right)\,\mathrm{Im}\, \tau}
  g_{mn}g_{pq}
  \bigl[\vp^{m},\vp^{p}\bigr]\bigl[\vp^{n},\vp^{q}\bigr].
\end{equation}
Using that the trace is cyclic we can write
\begin{equation}
  \partial_{m}B_{np}\,\tr\bigl\{\vp^{m}\bigl[\vp^{n},\vp^{p}\bigr]\bigr\}=
  \frac{2}{3}H_{mnp}\,\tr\bigl(\vp^{m}\vp^{n}\vp^{p}\bigr).
  \label{eq:DBI_3_point}
\end{equation}
Putting things together,
\begin{align}
  S_{\uD 3}^{\mathrm{DBI}}=-{\tau_{\uD 3}}\ell_{\us}^{4}
  \int\ud^{4}x\, \tr
  \biggl\{&\frac{\Phi_{+}+\Phi_{-}}{2\ell_{\us}^{4}}
  +\frac{\mathrm{Im}\, \tau}{4}
  f_{\mu\nu}f^{\mu\nu}
  +\frac{1}{2}g_{mn}D_{\mu}\varphi^{m}D^{\mu}\varphi ^{n}
    \notag\\
  &+\frac{1}{2\ell_{\us}^{2}}\partial_{m}\bigl(\Phi_{+}+\Phi_{-}\bigr)\vp^{m}
  +\frac{1}{4}\partial_{m}\partial_{n}\bigl(\Phi_{+}+\Phi_{-}\bigr)\vp^{m}
  \vp^{n}\notag\\
  &+\frac{\ui\left(\Phi_{+}+\Phi_{-}\right)}{24\, \mathrm{Im}\,\tau}
  \bigl(G_{+}-G_{-}+\overline{G}_{+}-\overline{G}_{-}\bigr)_{mnp}\vp^{m}
  \varphi^{n}\varphi^{p}\notag\\
  &-\frac{1}{4\, \mathrm{Im}\, \tau}\,g_{mn}g_{pq}\big[\varphi^{m},\varphi^{p}\bigr]
  \bigl[\varphi^{n},\varphi^{q}\bigr]\biggr\}.
  \label{eq:DBI_action}
\end{align}

Let's now turn to the CS action~\eqref{CS}.  In type-IIB, $n$ takes on
even values, $n=0,2,4,6,8$ where $C_{\left(6\right)}$ and
$C_{\left(8\right)}$ are the redundant magnetic duals of
$C_{\left(2\right)}$ and $C_{\left(0\right)}$ respectively.

For $n=0$, we write the contribution to the action as
\begin{multline}
  S_{\uD 3}^{0}=\pm\tau _{\uD 3}\int\Str\biggl\{
  \mathrm{P}\biggl[\biggl(1+\ui\ell_{\us}^{2}\iota_{\vp}^{2}
  -\frac{\ell_{\us}^{4}}{2}\iota_{\vp}^{4}\biggr) \biggl(
  C_{\left(0\right)}\bigl(\vp\bigr)\wedge
  \ue^{B_{\left(2\right)}\left(\vp\right)} \biggr) \biggr]\wedge
  \biggl(1+\ell_{\us}^{2} f_{\left(2\right)}+
  \frac{\ell_{\us}^{4}}{2}f_{\left(2\right)}\wedge
  f_{\left(2\right)}\biggr)\biggr\},
\end{multline}
where we have omitted terms that will only contribute at
$\cO\left(\ell_{\us}^{6}\right)$ or higher. The only terms that
contribute to the action are those that, after expanding
$\ue^{B_{\left(2\right)}}$, are $4$-forms.  Since $B_{\left(2\right)}$
has no legs on $R^{3,1}$, it contributes $\ell_{\us}^{4}$ from the
pullback and then another factor of $\ell_{\us}^{2}$ from the fact
that we chosen the gauge such the potential vanishes at the position
of the probe D3-branes. Thus terms in which $B_{\left(2\right)}$
contributes to ``soak up'' the legs of the integral are higher order
in $\ell_{\us}^{2}$ and so the $4$-form must be formed entirely from
$f_{\left(2\right)}\wedge f_{\left(2\right)}$.  Any scalars resulting
from the interior product acting on $\ue^{B_{\left(2\right)}}$ are
again higher order.  The result for the $n=0$ contribution through
$\cO\left(\ell_{\us}^{4}\right)$ is then
\begin{equation}
  S_{\uD 3}^{0}=\mp\frac{\ell_{\us}^{4}\tau_{\uD 3}}{8}\int \ud^{4}x\,\tr\biggl\{ 
  \mathrm{Re}\, \tau\, \ep ^{\mu \nu \rho \sigma}f_{\mu \nu}f_{\rho \sigma}\biggr\}.
\end{equation}

We can perform a similar argument for the $n=2$ contribution but since
$C_{\left(2\right)}$ has no legs on the non-compact directions, there
is no contribution through $\cO\left(\ell_{\us}^{4}\right)$.

For the $n=4$ contribution, we write
$C_{\left(4\right)}=C_{\left(4\right)}^{\,\mathrm{ext}}+C_{\left(4\right)}^{\,\mathrm{int}}$,
where from~\eqref{eq:GKP_ansatz}
$C_{\left(4\right)}^{\,\mathrm{ext}}=\alpha\,\ud\vol_{R^{3,1}}$ while
$C_{\left(4\right)}^{\,\mathrm{int}}$ has all four legs on the
internal manifold. Using the same reasoning as above, we find that
$C_{\left(4\right)}^{\,\mathrm{int}}$ does not contribute to action at
$\ell_{\us}^{4}$ order. For $C_{\left(4\right)}^{\,\mathrm{ext}}$, we
have
\begin{multline}
  S_{\uD 3}^{\,\mathrm{ext}}=\pm\tau _{\uD 3}\int\Str\biggl\{
  \mathrm{P}\biggl[\biggl(1+\ui\ell_{\us}^{2}\iota_{\vp}^{2}
  -\frac{\ell_{\us}^{4}}{2}\iota_{\vp}^{4}\biggr)
  \biggl(
  C_{\left(4\right)}^{\,\mathrm{ext}}\bigl(\vp\bigr)\wedge 
  \ue^{B_{\left(2\right)}\left(\vp\right)}\biggr)
  \biggr]\wedge \biggl(1+\ell_{\us}^{2} f_{\left(2\right)}+
  \frac{\ell_{\us}^{4}}{2}f_{\left(2\right)}
  \wedge f_{\left(2\right)}\biggr)\biggr\}.
\end{multline}
Now, since $\iota_{\vp}C_{\left(4\right)}^{\,\mathrm{ext}}=0$,
$C_{\left(4\right)}^{\,\mathrm{ext}}$ soaks up all of the legs and
this becomes
\begin{equation}
  S_{\uD 3}^{\,\mathrm{ext}}=\pm\tau_{\uD 3}
  \int\Str\biggl\{C_{\left(4\right)}^{\,\mathrm{ext}}\bigl(\vp\bigr)
  \biggl[1+\ui\ell_{\us}^{2}\iota_{\vp}^{2}B_{\left(2\right)}\bigl(\vp\bigr)
  -\frac{\ell_{\us}^{4}}{4}\iota_{\vp}^{4}\biggl(B_{\left(2\right)}\bigl(\vp\bigr)
  \wedge B_{\left(2\right)}\bigl(\vp\bigr)\biggr)\biggr]\biggr\}.
\end{equation}
We can make another gauge choice to set the constant part of
$C_{\left(4\right)}^{\,\mathrm{ext}}$ to zero, and so combining this
with the similar gauge choice for $B_{\left(2\right)}$, the Taylor
expansion gives
\begin{equation}
  S_{\uD 3}^{\,\mathrm{ext}}=\pm\tau_{\uD 3}\ell_{\us}^{4}
  \int\ud^{4}x\, \tr\biggl\{\frac{1}{2\ell_{\us}^{2}}
  \partial_{m}\bigl(\Phi_{+}-\Phi_{-}\bigr)\vp^{m}
  +\frac{1}{4}
  \partial_{m}\partial_{n}\bigl(\Phi_{+}-\Phi_{-}\bigr)\vp^{m}\vp^{n}
  \biggr\}.
\end{equation}

For $n=6$, the corresponding potential is defined by
\begin{equation}
  F_{\left(7\right)}=\ud C_{\left(6\right)}+C_{\left(4\right)}\wedge 
  H_{\left(3\right)}=-\hat{\ast}^{\left(\mathrm{s}\right)}F_{\left(3\right)},
\end{equation}
in which $\hat{\ast}^{\mathrm{\left(s\right)}}$ is the 10d Hodge-$\ast$ in the
string frame.  We find\footnote{Recall that our notation is that
  unadorned $\ast$ means the Hodge-$\ast$ built from the 6d unwarped
  metric $g_{mn}$.}
\begin{equation}
  F_{\left(7\right)}=-\ue^{4A+\phi}\ud\vol_{R^{3,1}}\wedge
  \ast F_{\left(3\right)},
\end{equation}
and thus $C_{\left(6\right)}$ has four legs on $R^{3,1}$ and two legs
on the internal space.  Setting the constant part of
$C_{\left(6\right)}$ to be a constant and applying reasoning similar
to the $C_{\left(4\right)}^{\,\mathrm{ext}}$ part gives the leading
order contribution
\begin{equation}
  S_{\uD 3}^{\mathrm{6}}=
  \pm\ui\tau_{\uD 3}\ell_{\us}^{2}
  \int\Str\biggl\{\iota_{\vp}^{2}C_{\left(6\right)}\bigl(\vp\bigr)\biggr\}.
\end{equation}
Writing
$C_{\left(6\right)}=\ud\vol_{R^{3,1}}\wedge\tilde{C}_{\left(2\right)}$,
the leading-order contribution is
\begin{equation}
  S_{\uD 3}^{\mathrm{6}}
  =\mp\frac{\ui\tau_{\uD 3}\ell_{\us}^{4}}{2}
  \int\ud^{4}x\, \tr\biggl\{\partial_{m}\tilde{C}_{np}\vp^{m}
  \bigl[\vp^{n},\vp^{p}\bigr]\biggr\}.
\end{equation}
Following the same steps that lead to~\eqref{eq:DBI_3_point} and using
$\ud\tilde{C}_{\left(2\right)}=-\ue^{4A+\phi}\ast F_{\left(3\right)}$
this term becomes
\begin{equation}
  S_{\uD 3}^{\mathrm{6}}
  =\pm\frac{\ui\tau_{\uD 3}\ell_{\us}^{4}}{24}
  \int\ud^{4}x\, \tr\biggl\{\frac{\Phi_{+}+\Phi_{-}}
  {\mathrm{Im}\, \tau}
  \bigl(G_{+}+G_{-}+\overline{G}_{+}+\overline{G}_{-}\bigr)_{mnp}
  \vp^{m}\vp^{n}\vp^{p}\biggr\}.
\end{equation}

Finally, we consider $n=8$ where the potential is defined via
\begin{equation}
  F_{\left(9\right)}=\ud C_{\left(8\right)}+C_{\left(6\right)}\wedge
  H_{\left(3\right)}=\hat{\ast}^{\left(\mathrm{s}\right)}F_{\left(1\right)}.
\end{equation}
Setting the constant part of $C_{\left(8\right)}$ to vanish, the
potential does not contribute to the action at this order as there is
a factor of $\ell_{\us}^{4}$ coming just from the interior product.

Combining these, we find
\begin{align}
  S_{\uD 3}^{\mathrm{CS}}=\pm\tau_{\uD 3}\ell_{\us}^{4}\int\ud^{4}x\,
  \tr\biggl\{&-\frac{\mathrm{Re}\,\tau}{8}
  \ep ^{\mu \nu \rho \sigma}f_{\mu \nu}f_{\rho \sigma}
  +\frac{1}{2\ell_{\us}^{2}}
  \partial_{m}\bigl(\Phi_{+}-\Phi_{-}\bigr)\vp^{m}\notag\\
  &+\frac{1}{4}
  \partial_{m}\partial_{n}\bigl(\Phi_{+}-\Phi_{-}\bigr)\vp^{m}\vp^{n}\notag\\
  &+\frac{\ui\left(\Phi_{+}+\Phi_{-}\right)}{24\, \mathrm{Im}\,\tau}
  \bigl(G_{+}+G_{-}+\overline{G}_{+}+\overline{G}_{-}\bigr)_{mnp}
  \vp^{m}\vp^{n}\vp^{p}\biggr\}.
  \label{eq:CS_action}
\end{align}

Adding this with~\eqref{eq:DBI_action}, we get the 4d Lagrangian for
the bosonic sector
\begin{align}
\label{eq:DBI_action_result}
  \cL^{\mathrm{B}}=\tr\biggl\{&
  -\frac{1}{4g^{2}}f_{\mu\nu}f^{\mu\nu}
  -\frac{\vartheta}{64\pi^{2}}\ep^{\mu\nu\rho\sigma}f_{\mu\nu}f_{\rho\sigma}
  -\frac{1}{2}K_{mn}D_{\mu}\vp^{m}D^{\mu}\vp^{n}\notag\\
  &-V_{0}-T_{m}\vp^{m}
  -\frac{1}{2}m^{2}_{\mathrm{B},mn}\vp^{m}\vp^{n}
  -\frac{\ui}{3!}C_{mnp}\vp^{m}\vp^{n}\vp^{p}
  +\frac{g^{2}}{4}K_{mn}K_{pq}
  \big[\varphi^{m},\varphi^{p}\bigr]
  \bigl[\varphi^{n},\varphi^{q}\bigr]\biggr\},
\end{align}
in which
\begin{subequations}
\begin{align}
  K_{mn}=&\frac{2\pi}{g_{\us}}g_{mn},\\
  g^{-2}&=\frac{2\pi}{g_{\us}}\mathrm{Im}\, \tau,\\
  \vartheta=&\pm\frac{16 \pi^{3}}{g_{\us}}\mathrm{Re}\,\tau,\\
  V_{0}=&\frac{\pi}{\ell_{\us}^{4}g_{\us}}\left(\Phi_{+}+\Phi_{-}\right),\\
  T_{m}=&\frac{2\pi}{\ell_{\us}^{2}g_{\us}}\partial_{m}\Phi_{\mp},\\
  m_{\mathrm{B},mn}^{2}=&\frac{2\pi}{g_{\us}}\partial_{m}\partial_{n}\Phi_{\mp},\\
  C_{mnp}=&\mp\frac{\pi}{g_{\us}}
  \frac{\Phi_{+}+\Phi_{-}}{\mathrm{Im}\,\tau}
  \bigl(G_{\mp}+\overline{G}_{\mp}\bigr)_{mnp},
\end{align}
\end{subequations}
where again the upper (lower) sign applies for $\uD 3$-branes
($\overline{\uD 3}$-branes).

\subsection{Fermionic action}

In this subsection, we consider the fermoinic modes on the D3.  We
begin with the Dirac-like action
of~\cite{Marolf:2003ye,*Marolf:2003vf, Martucci:2005rb} (see
also~\cite{Grana:2002tu}).  Although this action is applicable only in
the Abelian case of a single $\uD p$-brane, it is enough to deduce the
kinetic terms and mass terms.  The analogous action in the non-Abelian
case is not well-understood; however we will make use of a portion of
the action that follows from consistency with T-duality to determine
the Yukawa couplings.

\subsubsection{\label{sec:abelian_fermionic}Abelian case}

To leading order in $\ell_{\us}$, the fermionic action for a single
$\uD p$-brane in the Einstein frame is~\cite{Martucci:2005rb}
\begin{equation}
  \label{eq:abelian_Martucci}
  S^{\mathrm{F}}_{\uD p}=\ui \tau _{\uD p}\ell _{\us}^{4}
  \int\ud ^{p+1}\xi\, \ue^{\frac{p-3}{4}\phi}\sqrt{-\det\bigl(\hat{M}_{\alpha\beta}\bigr)}
  \bar{\Theta}P^{\uD p}_{\pm}\mathrm{P}
  \biggl[\bigl(\hat{\cM}^{-1}\bigr)^{\alpha\beta}
  \hat{\Ga} _{\beta}\biggl(\hat{\cD}_{\alpha}+\frac{1}{4}\hat{\Ga} _{\alpha}
  \hat{\cO} \biggl)-\hat{\cO}\biggr]\Theta,
\end{equation}
in which $\Theta$ is a double 10d Majorana-Weyl spinor (see
appendix~\ref{app:conv}) and again the upper (lower) sign applies to a
$\uD p$-brane ($\overline{\uD p}$-brane)\footnote{The sign difference
  in the projection operator with respect to~\cite{Martucci:2005rb} is
  a consequence of our different convention for the Levi-Civita
  tensor.}.  Note that in~\eqref{eq:abelian_Martucci} we have
redefined $\Theta$ with respect to~\cite{Martucci:2005rb} so that an
explicit power of $\ell_{\us}$ appears in order to match the one that
appears in the bosonic action~\eqref{eq:DBI_action_result}.
$\hat{M}_{\alpha\beta}$ is given by~\eqref{eq:generalized_metric}
(taken in the limit $\ell_{\us}\to 0$) while in IIB\footnote{In IIA,
  we make the replacement $\sig^{3}\to \II_{2}$.}
\begin{equation}
  \hat{\cM}_{\alpha\beta}
  =\mathrm{P}\bigl[\hat{g}_{\alpha\beta}\bigr]
  +\ue^{-\phi/2}\cF_{\alpha\beta}\Ga_{\left(10\right)}\otimes \sig^{3},
\end{equation}
where
\begin{equation}
  \cF_{\left(2\right)}=\mathrm{P}\bigl[B_{\left(2\right)}\bigr]
  +\ell_{\us}^{2}f_{\left(2\right)}.
\end{equation}
$\Ga_{\left(10\right)}$ is the 10d-chirality operator while $\sig^{3}$
acts on the extension space as discussed in
appendix~\ref{app:conv}.  The projection operator takes the form
\begin{subequations}
\label{eq:projection_operator}
\begin{equation}
  P^{\uD p}_{\pm}=
  \frac{1}{2}
  \begin{pmatrix} 1 & \pm \breve{\Ga}_{\uD p}^{-1} \\ 
    \pm \breve{\Ga}_{\uD p} & 1 \end{pmatrix},
\end{equation}
in which
\begin{align}
  \breve{\Ga}_{\uD p}=&\ui^{\left(p-2\right)\left(p-3\right)}
  \Ga_{\uD p}^{\left(0\right)}\Lambda\bigl(\cF\bigr),\\
  \Ga_{\uD p}^{\left(0\right)}=&\frac{1}{\left(p+1\right)!}
  \hat{\ve}_{\alpha_{1}\cdots\alpha_{p+1}}\hat{\Ga}^{\alpha_{1}\cdots\alpha_{p+1}},\\
  \La\bigl(\mathcal{F}\bigr)=&
  \frac{\sqrt{-\det \bigl(\mathrm{P}\bigl[\hat{g}_{\alpha\beta}\bigr]\bigr)}}
  {\sqrt{-\det\bigl(\hat{M}_{\alpha\beta}\bigr)}}
  \sum_{q}\frac{\ue^{-q\phi/2}}{2^{q}q!}
  \mathcal{F}_{\alpha _{1}\beta _{1}}\cdots{\cF}_{\alpha_{q}\beta_{q}}
  \hat{\Ga}^{\alpha_{1}\beta_{1}\cdots\alpha_{q}\beta_{q}}.
\end{align}
\end{subequations}
The operators $\hat{\cD}_{M}$ and $\hat{\cO}$ are related to the
supersymmetry transformations of the gravitino and
dilatino~\eqref{eq:susy_vars}.

For a $\uD 3$ probing~\eqref{eq:GKP_ansatz} with $\left\langle
  f_{\mu\nu}\right\rangle=0$, we have to leading order in $\ell_{\us}$
$\cF_{\left(2\right)}=0$ and $\hat{M}_{\mu\nu}=\ue^{2A}\eta_{\mu\nu}$.
This latter fact implies that $\hat{\cO}$ cancels out of the
action. Also to this order, only the leading term in $\Lambda
\left(\cF\right)$ contributes and so we take
$\Lambda\left(\cF\right)=1$, giving $\breve{\Ga}_{\uD
  3}=-\ui\Ga_{\left(4\right)}$.  Furthermore, in the
background~\eqref{eq:GKP_ansatz}, we have
\begin{equation}
  \hat{\cD}_{\mu}
  =\hat{\nabla}_{\mu}-\frac{1}{16}\ue^{\phi/2}
  \hat{\Ga}_{\mu}\hat{\cG}^{+}
  +\frac{1}{16}\hat{\slashed{F}}_{\left(5\right)}\hat{\Ga}_{\mu}
  \bigl(\ui \sig^{2}\bigr),
\end{equation}
where, as in~\eqref{eq:curly_G_2},
\begin{equation}
\label{eq:curly_G}
  \cG^{\pm}=\hat{\slashed{F}}_{\left(3\right)}\sig^{1}\pm 
  \ue^{-\phi}\hat{\slashed{H}}_{\left(3\right)}\sig^{3},
\end{equation}
and $\hat{\nabla}_{\mu}$ is the covariant derivative.  As is familiar
from the Green-Schwarz superstring, the fermionic
action~\eqref{eq:abelian_Martucci} is subject to a gauge redundancy
known as $\ka$-symmetry
\begin{equation}
  \Theta\sim\Theta+P_{\pm}^{\uD p}\ka,
\end{equation}
in which $\ka$ is an arbitrary double Majorana-Weyl spinor.  We can
use this to set
\begin{equation}
  \Theta=\begin{pmatrix} \theta \\ 0 \end{pmatrix},
\end{equation}
in which $\theta$ is an ordinary Majorana-Weyl spinor.  With this
choice of $\ka$-fixing, we find
\begin{align}
  S^{\mathrm{F}}_{\uD 3}=&\frac{\ui\tau_{\uD 3}\ell _{\us}^{4}}{2}
  \int\ud ^{4}x\, \ue^{4A}\bar{\theta}\biggl\{\hat{\Ga} ^{\mu}\hat{\nabla}_{\mu}
  \mp\frac{\ui}{16}
  \Ga_{\left(4\right)}\hat{\Ga}^{\mu}
  \hat{\slashed{F}}_{\left(5\right)}\hat{\Ga}_{\mu}
  -\frac{\ue^{\phi/2}}{4}\bigl(\pm \ui\Ga_{\left(4\right)}
  \hat{\slashed{F}}_{\left(3\right)}+\ue^{-\phi}\hat{\slashed{H}}_{\left(3\right)}
  \bigr)\biggr\}\theta .
\end{align}

From~\eqref{eq:spin_connection} we have
\begin{equation}
  \hat{\nabla}_{\mu}=\partial_{\mu}
  +\frac{1}{2}\partial_{m}A\, \hat{\Ga}_{\mu}\hat{\Ga}^{m}
  =\partial_{\mu}+\frac{\ue^{-2A}}{16}
  \partial_{m}\bigl(\Phi_{+}+\Phi_{-}\bigr)
  \bigl(\ga_{\mu}\ga_{\left(4\right)}\otimes \ga^{m}\bigr),
\end{equation}
where we have used the decomposition~\eqref{eq:10d_gamma_matrices} and
$\ga_{\mu}$ and $\ga_{m}$ are the unwarped $\ga$-matrices.  On the
other hand, from~\eqref{eq:GKP_ansatz} we have
\begin{equation}
  F_{\mu\nu\rho\sig m}=\ve_{\mu\nu\rho\sig}\partial_{m}\alpha,\qquad
  F_{mnpqr}=-\ue^{-8A}\ve_{mnpqr}^{\phantom{mnpqr}s}\partial_{s}\alpha,
\end{equation}
where $\ve_{123456}=\sqrt{\det\left(g_{mn}\right)}$ and similarly for
$\ve_{\mu\nu\rho\sig}$.  Hence,
\begin{equation}
  \hat{\slashed{F}}_{\left(5\right)}=-\ui \ue^{-3A}
  \partial_{m}\alpha\, \bigl(\II_{4}\otimes\ga^{m}\bigr)
  \bigl(1-\Ga_{\left(10\right)}\bigr).
\end{equation}
Using that $\Ga_{\left(10\right)}\theta=+\theta$, we find
\begin{equation}
  \hat{\slashed{F}}_{\left(5\right)}\hat{\Ga}_{\mu}\theta
  =-\ui\ue^{-2A}
  \partial_{m}\bigl(\Phi_{+}-\Phi_{-}\bigr)\bigl(\ga_{\mu}\otimes \ga^{m}\bigr)
  \theta.
\end{equation}
Thus, the action becomes
\begin{align}
  S_{\uD 3}^{\mathrm{F}}
  =\frac{\ui\tau_{\uD 3}\ell_{\us}^{4}}{2}
  \int\ud^{4}x\,
  \bar{\theta}
  \biggl\{&\ue^{3A}
  \slashed{\partial}\otimes \II_{8}
  +\frac{\ue^{A}}{2}\partial_{m}\Phi_{\mp}
  \ga_{\left(4\right)}\otimes\ga^{m}\notag\\
  &\mp\frac{\ui\ue^{7A+\phi/2}}{8}\bigl[
  \bigl(\II_{4}\mp\ga_{\left(4\right)}\bigr)\otimes
  \slashed{G}_{\left(3\right)}
  +\bigl(\II_{4}\pm\ga_{\left(4\right)}\bigr)\otimes
  \overline{\slashed{G}}_{\left(3\right)}\bigr]\biggr\}\theta,
\end{align}
where, for example,
$\slashed{G}_{\left(3\right)}=\frac{1}{3!}G_{mnp}{\ga}^{mnp}$ involves
only unwarped $\SO{6}$ $\ga$-matrices and similarly
$\slashed{\partial}=\ga^{\mu}\partial_{\mu}$.  If $\eta_{\pm}$ is a 6d
Weyl spinor satisfying $\ga_{\left(6\right)}\eta_{\pm}=\pm\eta_{\pm}$,
then
\begin{equation}
  \ga^{mnp}\eta_{\pm}=\pm\frac{\ui}{3!}\ep^{mnp}_{\phantom{mnp}stl}
  \ga^{stl}\eta_{\pm},
\end{equation}
and so for a $3$-form $X_{\left(3\right)}$,
\begin{equation}
\label{eq:slash_Hodge}
  \slashed{X}_{\left(3\right)}\eta_{\pm}
  =\mp \ui \slashed{\widetilde{X}}_{\left(3\right)}\eta_{\pm},
\end{equation}
in which $\widetilde{X}_{\left(3\right)}=\ast X_{\left(3\right)}$.
Since $\Ga_{\left(10\right)}\theta=+\theta$, we have
$\Ga_{\left(4\right)}\theta=\Ga_{\left(6\right)}\theta$ and so
\begin{align}
\label{eq:D3_partial_action}
  S_{\uD 3}^{\mathrm{F}}
  =\frac{\ui\tau_{\uD 3}\ell_{\us}^{4}}{2}
  \int\ud^{4}x\,
  \bar{\theta}
  \biggl\{&\ue^{3A}
  \slashed{\partial}\otimes \II_{8}
  +\frac{\ue^{A}}{2}
  \ga_{\left(4\right)}\otimes\slashed{\partial}\Phi_{\mp}\notag\\
  &+\frac{\ue^{7A+\phi/2}}{16}\bigl[
  \bigl(\II_{4}\mp\ga_{\left(4\right)}\bigr)\otimes
  \slashed{G}_{\mp}
  -\bigl(\II_{4}\pm\ga_{\left(4\right)}\bigr)\otimes
  \overline{\slashed{G}}_{\mp}\bigr]\biggr\}\theta.
\end{align}

The fermionic modes on the $\uD 3$ can be decomposed into a gaugino
$\lambda$ and a number of modulini $\psi^{m}$, the fermionic partners of the
transverse deformations of the worldvolume,
\begin{equation}
  \theta = \theta_{\mathrm{g}}+\theta_{\mathrm{m}}.
\end{equation}
Following~\cite{Grana:2003ek}, we can determine how to extract these
modes by considering the supersymmetry transformations.  To this end,
we consider the case where the metric and fluxes satisfy the
conditions for $\cN_{4}=1$ supersymmetry.  Then the solution to the
Killing spinor equations
\begin{equation}
  \hat{\cD}_{M}\hat{\ep}=0,\qquad \hat{\cO}\hat{\ep}=0,
\end{equation}
takes the form
\begin{equation}
  \hat{\ep}=\begin{pmatrix} \hat{\ep}_{1} \\ \hat{\ep}_{2} \end{pmatrix},
\end{equation}
where~\cite{Grana:2001xn}
\begin{align}
  \hat{\ep}_{1}=&\ue^{A/2}\begin{pmatrix} 0 \\ \ep_{\alpha}\end{pmatrix}\otimes
  \eta_{-}
  -\ue^{A/2}\begin{pmatrix} \ui\,\bar{\ep}^{\dot{\alpha}} \\ 0 \end{pmatrix}
  \otimes \eta_{+},\notag\\
  \hat{\ep}_{2}=&-\ui\,\ue^{A/2}\begin{pmatrix} 0 \\ \ep_{\alpha}\end{pmatrix}\otimes
  \eta_{-}
  -\ui\,\ue^{A/2}\begin{pmatrix} \ui\,\bar{\ep}^{\dot{\alpha}} \\ 0 \end{pmatrix}
  \otimes \eta_{+},
\end{align}
in which $\ep_{\alpha}$ is an arbitrary constant spinor, $\eta_{-}$ is a
negative chirality spinor satisfying
\begin{equation}
\label{eq:Killing_spinor}
  0=\nabla_{m}\eta_{-}+\frac{\ui}{4}\ue^{\phi}F_{m}\eta_{-},
\end{equation}
and $\eta_{+}:=B_{6}^{\ast}\eta_{-}^{\ast}$.  In the string frame the
supersymmetry transformations of the $\uD 3$ bosonic fields take
the schematic forms
\begin{equation}
  \delta_{\ep}A_{\mu}\sim \bar{\Theta}^{\left(\mathrm{s}\right)}
  \hat{\Ga}_{\mu}^{\left(\mathrm{s}\right)}
  \hat{\ep}^{\left(\mathrm{s}\right)},\qquad
  \delta_{\ep}{\Phi}^{m}\sim \bar{\Theta}^{\left(\mathrm{s}\right)}
  \hat{\Ga}^{m\left(\mathrm{s}\right)}\hat{\ep}^{\left(\mathrm{s}\right)}
\end{equation}
Moving to the Einstein frame,
$\hat{g}_{MN}=e^{-\phi/2}\hat{g}_{MN}^{\left(\mathrm{s}\right)}$,
$\hat{\ep}=\ue^{-\phi/8}\hat{\ep}^{\left(\mathrm{s}\right)}$,
${\Theta}=\ue^{-\phi/8}\Theta^{\left(\mathrm{s}\right)}$,
we have
\begin{equation}
\label{eq:D3_susy_var_einstein}
  \delta_{\ep}A_{\mu}\sim \ue^{\phi/2}\bar{\Theta}\hat{\Ga}_{\mu}\hat{\ep},
  \qquad
  \delta_{\ep}{\Phi}^{m}\sim \bar{\Theta}
  \hat{\Ga}^{m}\hat{\ep}.
\end{equation}
We wish to recover the usual $\cN_{4}=1$ supersymmetry transformations
\begin{equation}
\label{eq:4d_susy_var}
  \delta A_{\mu}\sim \lambda\ga_{\mu}\ep,\qquad
  \delta \vp^{i}\sim \psi^{i}\ep,
\end{equation}
where we have used $\eta_{-}$ to define a complex structure
characterized by the $\left(3,0\right)$ form
\begin{equation}
  \Omega_{mnp}=\eta_{-}^{\dagger}\ga_{mnp}\eta_{+},
\end{equation}
and have denoted holomorphic and anti-holomorphic indices by $i$ and
$\bar{\imath}$.  Then~\eqref{eq:4d_susy_var} is recovered
from~\eqref{eq:D3_susy_var_einstein} by taking
\begin{align}
  \theta_{\mathrm{g}}=&a\,\ue^{-3A/2-\phi/2}
  \begin{pmatrix} 0 \\ \lambda_{\alpha}\end{pmatrix}
  \otimes\eta_{-}
  -a\,\ue^{-3A/2-\phi/2}
  \begin{pmatrix}
    \ui\,\bar{\lambda}^{\dot{\alpha}} \\ 0
  \end{pmatrix}
  \otimes\eta_{+},\notag\\
  \theta_{\mathrm{m}}=&b\, \ue^{-3A/2}
  \begin{pmatrix}
    0 \\ \psi^{i}_{\alpha}
  \end{pmatrix}
  \otimes \Omega_{ijk}\ga^{jk}\eta_{-}
  -b\,\ue^{-3A/2}
  \begin{pmatrix}
    \ui\,\bar{\psi}^{\bar{\imath}\dot{\alpha}} \\ 0
  \end{pmatrix}
  \bar{\Omega}_{\bar{\imath}\bar{\jmath}\bar{k}}\ga^{\bar{\jmath}\bar{k}}\eta_{+},
\end{align}
in which $a$ and $b$ are normalization constants.  Note that the form
is taken to ensure that $\theta$ is Majorana-Weyl.

Consider now the non-supersymmetric case.  As discussed in the
previous section, generically, the addition of $\overline{\uD
  3}$-branes will cause the metric to no longer be Hermitian with
respect to the complex structure.  However, at least away from the
$\overline{\uD 3}$, the spinor $\eta_{-}$ defines an $\SU{3}$
structure\footnote{See, e.g.,~\cite{Grana:2005jc, *Koerber:2010bx} for
  reviews on $G$-structures.} and from this we can construct an almost
complex structure $J_{m}^{\phantom{m}n}$ and a pre-symplectic
structure $\omega_{mn}$.  The existence of the former is equivalent to
the existence of a $3$-form $\Omega$ and we have
\begin{equation}
\label{eq:structure}
  \Omega_{mnp}=\eta_{-}^{\dagger}\ga_{mnp}\eta_{+},\qquad
  \omega_{mn}=\ui \eta_{+}^{\dagger}\ga_{mn}\eta_{+}.
\end{equation}
We emphasize that, since in the non-supersymmetric case there is no
natural spinor to define them, these structures are defined by the
spinor satisfying~\eqref{eq:Killing_spinor} where the derivative is
built from the unperturbed K\"ahler metric of the supersymmetric
solution that we are perturbing.  We also note that we are no longer
guaranteed that $\eta_{-}$ is well-defined and non-vanishing
everywhere in the internal space, and so these structures may only be
locally defined. By construction, these structures satisfy the
compatibility condition
\begin{equation}
  \label{eq:compatibility}
  \Omega\wedge\omega=0,
\end{equation}
which ensures that the metric that defines the Clifford algebra is
Hermitian and we have $\omega_{mn}=J_{mn}$.  However, in general we
are not ensured that either $\omega$ nor $\Omega$ is closed and so the
space is not immediately K\"ahler or indeed even complex.  Therefore, 
we will not, for now, explicitly denote indices that are holomorphic
or anti-holomorphic with respect to this perturbed almost complex
structure and  write
\begin{equation}
  \theta_{\mathrm{m}}=b\, \ue^{-3A/2}
  \begin{pmatrix}
    0 \\ \psi^{m}_{\alpha}
  \end{pmatrix}
  \otimes \Omega_{mnp}\ga^{np}\eta_{-}
  -b\,\ue^{-3A/2}
  \begin{pmatrix}
    \ui\,\bar{\psi}^{m\dot{\alpha}} \\ 0
  \end{pmatrix}\otimes
  {\Omega}^{\ast}_{mnp}\ga^{np}\eta_{+}.
\end{equation}
Note that although the notation suggests that there are now six
independent Weyl fermions in 4d, the fact that $\eta_{-}$ is Weyl, and
therefore pure in the sense that it is annihilated by half of the
$\ga$-matrices, implies that only three of them are independent.  In
the supersymmetric case, the analogous statement is
$\psi^{\bar{\imath}}=0$ (since $\Omega_{\bar{\imath}mn}=0$) where
$\psi^{\bar{\imath}}$ should not be confused with
$\bar{\psi}^{\bar{{\imath}}}=\bigl(\psi^{i}\bigr)^{\ast}$.

Consider now~\eqref{eq:D3_partial_action}.  The first operator that
appears gives rise to the 4d kinetic terms and we have
\begin{equation}
  \ue^{3A}\bar{\theta}_{\mathrm{g}}
  \,\slashed{\partial}\otimes\II_{8}\,\theta_{\mathrm{g}}
  =-a^{2}\ue^{-\phi}
  \biggl\{\bar{\lambda}\bar{\sig}^{\mu}\partial_{\mu}\lambda\,
  \eta_{-}^{\dagger}\eta_{-}+\lambda\sig^{\mu}\partial_{\mu}\bar{\lambda}\,
  \eta_{+}^{\dagger}\eta_{+}\biggr\}.
\end{equation}
We normalize $\eta_{-}$ so that at the position of the $\uD 3$,
\begin{equation}
  \eta_{-}^{\dagger}\eta_{-}=\eta_{+}^{\dagger}\eta_{+}=1.
\end{equation}
The factor of $\ue^{-\phi}$ is what is expected from the kinetic term
of $A_{\mu}$ appearing in~\eqref{eq:DBI_action_result} and so we get a
properly normalized term by setting $a=1$.

Next, we consider
\begin{equation}
  \ue^{3A}\bar{\theta}_{\mathrm{g}}\,\slashed{\partial}\otimes\II_{8}\,
  \theta_{\mathrm{m}}
  =-ab\,\ue^{-\phi/2}\biggl\{\bar{\lam}\bar{\sig}^{\mu}\partial_{\mu}\psi^{m}
  \, \Omega_{mnp}\,\eta_{-}^{\dagger}\ga^{np}\eta_{-}
  +\lam\sig^{\mu}\partial_{\mu}\bar{\psi}^{m}\,
  \Omega^{\ast}_{mnp}
  \,\eta_{+}^{\dagger}\ga^{np}\eta_{+}\biggr\}.
\end{equation}
Using~\eqref{eq:structure}, we see that these terms depend on
$\Omega_{mnp}\,\omega^{np}$ which vanishes as a consequence of
compatibility~\eqref{eq:compatibility} and the fact that $\Omega$ is
IASD (using~\eqref{eq:structure} and~\eqref{eq:slash_Hodge}).  We then
have
\begin{equation}
  \ue^{3A}\bar{\theta}_{\mathrm{g}}\,\slashed{\partial}\otimes\II_{8}\,
  \theta_{\mathrm{m}}
  =\ue^{3A}\bar{\theta}_{\mathrm{m}}\,\slashed{\partial}\otimes\II_{8}\,
  \theta_{\mathrm{g}}=0.
\end{equation}

The last kinetic term is
\begin{multline}
  \ue^{3A}\bar{\theta}_{\mathrm{m}}\,\slashed{\partial}\otimes\II_{8}\,
  \theta_{\mathrm{m}}
  \\=-b^{2}\,\biggl\{\bar{\psi}^{m}\bar{\sig}^{\mu}\partial_{\mu}\psi^{n}
  \, \Omega^{\ast}_{mpq}\Omega_{nst}\,\eta_{-}^{\dagger}\ga^{qp}\ga^{st}\eta_{-}
  +\psi^{m}\sig^{\mu}\partial_{\mu}\bar{\psi}^{n}\,
  \Omega_{mpq}\Omega^{\ast}_{nst}
  \,\eta_{+}^{\dagger}\ga^{qp}\ga^{st}\eta_{+}\biggr\}.
\end{multline}
Making use of the Clifford algebra, the fact that
$\ast\omega=\frac{1}{2}\omega\wedge\omega$, the compatibility of the
almost complex and pre-symplectic structures, and the identity
\begin{equation}
  \ga^{mnpq}\eta_{\pm}=
  \mp\frac{\ui}{2!}\ep^{mnpq}_{\phantom{mnpq}st}\ga^{st}\eta_{\pm},
\end{equation}
we have
\begin{equation}
\label{eq:Omega_contraction}
  \Omega^{\ast}_{mpq}\Omega_{nst}\,\eta_{-}^{\dagger}\ga^{qp}\ga^{st}\eta_{-}=
  8\Omega_{m}^{\ast\ pq}\Omega_{npq}
  ={8}\left\lvert\Omega\right\rvert^{2}\bigl(g_{mn}-\ui\,\omega_{mn}\bigr),
\end{equation}
where we use the notation~\eqref{eq:define_dot_product}.  Thus,
setting
\begin{equation}
  b=\frac{1}{4\left\vert\Omega\right\vert},
\end{equation}
we get
\begin{equation}
  \ue^{3A}\bar{\theta}_{\mathrm{m}}\,\slashed{\partial}\otimes\II_{8}\,
  \theta_{\mathrm{m}}\\
  =-\frac{1}{2}\bigl(g_{mn}-\ui\,\omega_{mn}\bigr)
  \,\bar{\psi}^{m}\bar{\sig}^{\mu}\partial_{\mu}\psi^{n}
  -\frac{1}{2}\bigl(g_{mn}+\ui\,\omega_{mn}\bigr)
  \,\psi^{m}\sig^{\mu}\partial_{\mu}\bar{\psi}^{n}.
\end{equation}
Summarizing, after integrating by parts the kinetic terms are
\begin{equation}
  -\ui\tau_{\uD 3}\ell_{\us}^{4}\int\ud^{4}x\,
  \bigl\{\mathrm{Im}\, \tau\, \bar{\lam}\bar{\sig}^{\mu}\partial_{\mu}\lam
  +\frac{1}{2}\bigl(g_{mn}-\ui\,\omega_{mn}\bigr)
  \bar{\psi}^{m}\bar{\sig}^{\mu}\partial_{\mu}\psi^{n}\bigr\}.
\end{equation}

The next operator in~\eqref{eq:D3_partial_action} is the coupling to
$\Phi_{\pm}$.  However, since $\theta$ is Majorana-Weyl, any bilinear
of the type
\begin{equation}
  \bar{\theta}\hat{\Ga}^{M_{1}\cdots M_{n}}\theta,
\end{equation}
automatically vanishes unless $n$ is $3$ or $7$ and hence this
coupling vanishes.

The masses therefore come only from the $3$-form contribution.  For
the D3 case,
\begin{equation}
  S_{\uD 3}^{\, 3}=
  \frac{\ui\tau_{\uD 3}\ell_{\us}^{4}}{32}\int\ud^{4}x\,\ue^{7A+\phi/2}
  \bar{\theta}\bigl\{\bigl(\II_{4}-\ga_{\left(4\right)}\bigr)
  \otimes\slashed{G}_{-}
  -\bigl(\II_{4}+\ga_{\left(4\right)}\bigr)\otimes
  \overline{\slashed{G}}_{-}
  \bigr\}\theta.
\end{equation}
Consider
\begin{equation}
  \frac{\ue^{7A+\phi/2}}{32}\bar{\theta}_{\mathrm{g}}
  \bigl(\II_{4}-\ga_{\left(4\right)}\bigr)\otimes\slashed{G}_{-}
  \theta_{\mathrm{g}}
  =-\frac{\ui\ue^{4A-\phi/2}a^{2}}{16}\lam\lam\,
  \eta_{+}^{\dagger}\slashed{G}_{-}\eta_{-}.
\end{equation}
From~\eqref{eq:structure}, we have
\begin{equation}
  \eta_{+}^{\dagger}\ga_{mnp}\eta_{-}=-\Omega^{\ast}_{mnp},
\end{equation}
so this becomes
\begin{equation}
  \frac{\ui\ue^{4A-\phi/2}a^{2}}{16}
  G_{-}\cdot\overline{\Omega}\,\lam\lam,
\end{equation}
where again we recall~\eqref{eq:define_dot_product}.  Note that if the
complex structure were not perturbed this would  provide a
coupling to the $\left(3,0\right)$ part of $G_{\left(3\right)}$ alone.
However, in general this will couple also to other (unperturbed)
Hodge-types. The term in the action is
\begin{equation}
  -\tau_{\uD 3}{\ell_{\us}^{4}}
  \int\ud^{4}x\, \frac{\left(\Phi_{+}+\Phi_{-}\right)
    \left(\mathrm{Im}\, \tau\right)^{1/2}}{32}
  \biggl\{G_{-}\cdot\overline{\Omega}\, \lambda\lambda
  + \overline{G}_{-}\cdot\Omega\, \bar{\lambda}\bar{\lambda}\biggr\}.
\end{equation}

Next, we consider the terms that mix the gaugino and the modulini in
the mass matrix
\begin{equation}
  \frac{\ue^{7A+\phi/2}}{32}\bar{\theta}_{\mathrm{g}}
  \bigl(\II_{4}-\ga_{\left(4\right)}\bigr)\otimes
  \slashed{G}_{-}\theta_{\mathrm{m}}
  =-\frac{\ui\ue^{4A}ab}{16}
  \lam\psi^{m}\, \eta_{+}^{\dagger}\slashed{G}_{-}\Omega_{mnp}
  \ga^{np}\eta_{-}.
\end{equation}
One can show $\eta_{+}^{\dagger}\ga_{m}\eta_{-}=0$ which implies that
$\eta_{+}^{\dagger}\ga_{mnpqr}\eta_{-}=0$  and hence, using the Clifford
algebra, we find
\begin{equation}
  \eta_{+}^{\dagger}\slashed{G}_{-}\Omega_{mnp}
  \ga^{np}\eta_{-}=
  -\Omega_{mnp}\Omega_{tl}^{\ast\ p}
  G_{-}^{ntl}.
\end{equation}
Using
\begin{equation}
  \Omega_{tl}^{\ast\ p}\Omega_{mnp}
  =\frac{\left\lvert\Omega\right\rvert^{2}}{4}\biggl[
  \bigl(g_{tm}-\ui\,\omega_{tm}\bigr)\bigl(g_{ln}-\ui\,\omega_{ln}\bigr)
  -
  \bigl(g_{lm}-\ui\,\omega_{lm}\bigr)\bigl(g_{tn}-\ui\,\omega_{tn}\bigr)\biggr],
\end{equation}
we find
\begin{equation}
  \eta_{+}^{\dagger}\slashed{G}_{-}\Omega_{mnp}
  \ga^{np}\eta_{-}=
  \frac{\ui\left\lvert\Omega\right\rvert^{2}}{2}
  \bigl(g_{ml}+\ui\,\omega_{ml}\bigr)G_{-nt}^{l}\omega^{nt}.
\end{equation}
Hence, this coupling corresponds to the non-primitive part of the
$\left(2,1\right)$-flux in the case in which the complex structure is
not perturbed\footnote{Recall however that generic Calabi-Yaus and
  other simply connected spaces have $b_{1}=b_{5}=0$ (where $b_{i}$
  are the Betti numbers) and so do not support non-primitive flux
  since $\omega\wedge X_{\left(3\right)}=0$ automatically.}.  We get
the same result coming from
$\bar{\theta}_{\mathrm{m}}\bigl(\II_{4}-\ga_{\left(4\right)}\bigr)\otimes\slashed{G}_{-}\theta_{\mathrm{m}}$
and hence the gaugino-modulino mass-mixing is
\begin{equation}
  \ui\tau_{\uD 3}\ell_{\us}^{4}\int\ud^{4}x\,
  \frac{\left(\Phi_{+}+\Phi_{-}\right)
    \left\lvert\Omega\right\rvert}
  {128}
  \biggl\{\lam\psi^{m}\bigl(g_{ml}+\ui\,\omega_{ml}\bigr)
  G^{l}_{-nt}\omega^{nt}
  -
  \bar{\lam}\bar{\psi}^{m}\bigl(g_{ml}-\ui\,\omega_{ml}\bigr)
  \overline{G}_{-nt}^{l}\omega^{nt}\biggr\}.
\end{equation}

The final contribution to the mass matrix is the modulino-modulino
part.  We have
\begin{equation}
  \frac{\ue^{7A+\phi/2}}{32}
  \bar{\theta}_{\mathrm{m}}
  \bigl(\II_{4}-\ga_{\left(4\right)}\bigr)\otimes\slashed{G}_{-}
  \theta_{\mathrm{m}}
  =-\frac{\ui\ue^{4A+\phi/2}b^{2}}{16}
  \psi^{m}\psi^{n}\eta_{+}^{\dagger}\Omega_{mpq}\ga^{qp}\slashed{G}_{-}
  \Omega_{nst}\ga^{st}\eta_{-}.
\end{equation}
Since $\psi^{m}\psi^{n}$ is symmetric in $m$ and $n$ this becomes
\begin{equation}
  \psi^{m}\psi^{n}\eta_{+}^{\dagger}\Omega_{mpq}\ga^{qp}\slashed{G}_{-}
  \Omega_{nst}\ga^{st}\eta_{-}
  =-4\bigl\lvert\Omega\bigr\rvert^{2}
  \bigl(g_{l\left(m\right.}-\ui\,\omega_{l\left(m\right.}\bigr)
  \Omega_{\left.n\right)pq}G_{-}^{lpq}
  \psi ^{m}\psi ^{n},
\end{equation}
and so the corresponding part of the action is
\begin{align}
  -\tau_{\uD 3}\ell_{\us}^{4}\int\ud^{4}x\,
  \frac{\Phi_{+}+\Phi_{-}}
  {128\left(\mathrm{Im}\,\tau\right)^{1/2}}
  \biggl\{&\psi^{m}\psi^{n}
  \bigl[g_{l\left(m\right.}-\ui\,\omega_{l\left(m\right.}\bigr]
  \Omega_{\left.n\right)pq}G_{-}^{lpq}\notag\\
  +&
  \bar{\psi}^{m}\bar{\psi}^{n}
  \bigl[g_{l\left(m\right.}+\ui\,\omega_{l\left(m\right.}\bigr]
  \bar{\Omega}_{\left.n\right)pq}\overline{G}_{-}^{lpq}
  \biggr\}.
\end{align}
This couples to the primitive $\left(1,2\right)$ flux in the case in
which the complex structure is not perturbed.

\subsubsection{\label{sec:fermionic_nonabelian}Non-Abelian case}

The previous analysis in the Abelian case suffices to determine the
kinetic terms and mass terms in the fermionic action, but in order to
determine the Yukawa couplings we need to move to the non-Abelian
case.  Unfortunately, the non-Abelian version of the fermionic
action~\eqref{eq:abelian_Martucci} is not currently well-understood.
However, we can argue from T-duality and supersymmetry how the
action~\eqref{eq:abelian_Martucci} will be modified to leading order
in $\ell_{\us}$ for our backgrounds of interest\footnote{In addition
  to the term that we consider here, there may be
  $\ell_{\us}$-suppressed Yukawa couplings arising from, for example,
  performing a Taylor expansion of closed-string fields.}.

To do so, we again consider a stack of $N$ $\uD p$-branes.  As
discussed in section~\ref{sec:bosonic}, this involves the promotion of
the gauge symmetry to $\U{N}$ and the corresponding modification to
the connection $A_{\left(1\right)}$ and its curvature.  This of course
must be accompanied by the modification of the ordinary derivative to
the gauge-covariant derivative.  However, since we have taken
$\left\langle f_{\left(2\right)}\right\rangle=0$, there are no other
modifications to marginal or relevant operators from this modification
to $A_{\left(1\right)}$.  A further change is the promotion of the
transverse fluctuations to adjoint-valued fields and the Taylor
expansion to non-Abelian Taylor expansions.  But, as even the usual
Taylor expansion in~\eqref{eq:abelian_Martucci} will lead only to
$\ell_{\us}$-corrections, this again will not be relevant.  The
fermionic variables themselves are promoted to adjoint-valued fields
and the non-Abelian action must contain a trace that is symmetrized
according to some procedure.  Fortunately, since all of the operators
discussed in the previous section are quadratic in the fermions and,
to this order in $\ell_{\us}$, the closed string fields are
proportional to the identity, this becomes a simple trace.  As a
result, part of the action is (c.f.~\eqref{eq:D3_partial_action})
\begin{align}
\label{eq:D3_na_partial_action_a}
  S_{\uD 3}^{\mathrm{F}}
  \ni\frac{\ui\tau_{\uD 3}\ell_{\us}^{4}}{2}
  \int\ud^{4}x\,\tr\biggl[
  \bar{\theta}
  \biggl\{&\ue^{3A}
  \slashed{\partial}\otimes \II_{8}
  +\frac{\ue^{A}}{2}
  \ga_{\left(4\right)}\otimes\slashed{\partial}\Phi_{\mp}\notag\\
  &+\frac{\ue^{7A+\phi/2}}{16}\bigl[
  \bigl(\II_{4}\mp\ga_{\left(4\right)}\bigr)\otimes
  \slashed{G}_{\mp}
  -\bigl(\II_{4}\pm\ga_{\left(4\right)}\bigr)\otimes
  \overline{\slashed{G}}_{\mp}\bigr]\biggr\}\theta\biggr].
\end{align}

A further modification, required by gauge invariance, is the
replacement of the ordinary derivative $\partial_{\mu}$ with the
gauge-covariant derivative
\begin{equation}
  D_{\alpha}=\partial_{\alpha}-\ui\left[A_{\alpha},\right].
\end{equation}
This leads to an additional term in the action and in the absence of
fluxes, the $\ka$-fixed string-frame action includes, for any $p$
\begin{equation}
  \label{0yukawa}
  \frac{\tau _{\uD p}\ell _{\us}^{4}}{2}
  \int \ud^{p+1}\xi\, \ue^{-\phi}\tr 
  \biggl\{\bar{\theta}^{\left(\mathrm{s}\right)}\,
  \hat{\Ga}^{\left(s\right)\alpha}
  \bigl[A_{\alpha},\theta^{\left(s\right)}\bigr]\biggr\}.
\end{equation}
For this generalization to be consistent with T-duality under which
$A_{\alpha}$ is exchanged with transverse deformations $\vp^{i}$, we
must include the term
\begin{equation}
  \frac{\tau _{\uD p}\ell_{\us}^{4}}{2}\int \ud^{p+1}\xi\, \ue^{-\phi}
  \tr\biggl\{\bar{\theta}^{\left(\mathrm{s}\right)}
  \hat{\Ga}^{\left(\mathrm{s}\right)}_{i}
  \bigl[\varphi^{i},\theta^{\left(\mathrm{s}\right)}\bigr]\biggr\}.
\end{equation}
We can confirm that at this level no symmetrization prescription is
required since these couplings agree with the expectation from
supersymmetry (see also~\cite{Wynants}).  In the presence of fluxes,
it is natural, given the bosonic action~\eqref{eq:Myers}, to expect
that the worldvolume indices ought to be contracted with
$\hat{M}_{\alpha\beta}$ (or $\hat{\cM}_{\alpha\beta}$ before
$\ka$-fixing) while transverse indices ought to be contracted with
$\hat{E}_{mn}$ and its inverse.  However, taking the gauge choice
$B_{\left(2\right)}=0$ at the position of the $\uD 3$s, these effects
do not contribute at this order in $\ell_{\us}$.

In summary, to leading order in $\ell_{\us}$, the effect of moving to
the non-Abelian case in our background is to
replace~\eqref{eq:D3_partial_action} with
\begin{align}
\label{eq:D3_na_partial_action}
  S_{\uD 3}^{\mathrm{F}}
  =\frac{\ui\tau_{\uD 3}\ell_{\us}^{4}}{2}
  \int\ud^{4}x\,\tr\biggl[
  \bar{\theta}
  \biggl\{&\ue^{3A}
  \slashed{D}\otimes \II_{8}
  +\frac{\ue^{A}}{2}
  \ga_{\left(4\right)}\otimes\slashed{\partial}\Phi_{\mp}
  -\ui\,\ue^{3A+\phi/2}\bigl(\ga_{\left(4\right)}\otimes\ga_{m}\bigr)
  \bigl[\vp^{m},\cdot\bigr]\notag\\
  &+\frac{\ue^{7A+\phi/2}}{16}\bigl[
  \bigl(\II_{4}\mp\ga_{\left(4\right)}\bigr)\otimes
  \slashed{G}_{\mp}
  -\bigl(\II_{4}\pm\ga_{\left(4\right)}\bigr)\otimes
  \overline{\slashed{G}}_{\mp}\bigr]\biggr\}\theta\biggr].
\end{align}
The factors of $\ue^{\phi}$ arise from moving to the 10d Einstein
frame.  The same sign for the Yukawa applies for both the
${\uD 3}$ case and the $\overline{\uD 3}$ case since it
results from the supersymmetrization of the DBI part of the bosonic action
which is independent of the sign of the $\uD 3$-brane charge.

For the kinetic and mass terms, the modification from the Abelian case
is minimal since, in our normalization, the generators satisfy
$\tr\bigl(T^{a}T^{b}\bigr)=\delta^{ab}$.  However, the Yukawa
couplings involve further analysis.  Note that because of the
non-trivial gauge structure, the bilinear doesn't automatically vanish
even though there is only a single $\hat{\Ga}$-matrix present
($\hat{\Ga}_{m}=\ue^{-A}\ga_{\left(4\right)}\otimes \ga_{m}$).
However, for the term arising when $\theta$ is pure gaugino, we have
\begin{multline}
  -\ui\ue^{3A+\phi/2}\,\tr\,\biggl\{\bar{\theta}_{\mathrm{g}}\bigl(
  \ga_{\left(4\right)}\otimes\ga_{m}\bigr)
  \bigl[\vp^{m},\theta_{\mathrm{g}}\bigr]\biggr\}\\
  =a^{2}\ue^{-\phi/2}\tr\biggl\{\lam\bigl[\vp^{m},\lam\bigr]\biggr\}
  \eta_{+}^{\dagger}\ga_{m}\eta_{-}
  -a^{2}\ue^{-\phi/2}\tr\biggl\{\bar{\lam}\bigl[\vp^{m},\bar{\lam}\bigr]\biggr\}
  \eta_{-}^{\dagger}\ga_{m}\eta_{+},
\end{multline}
which, on account of the fact that
$\eta_{+}^{\dagger}\ga_{m}\eta_{-}=0$, does vanish.

For terms involving the gaugino and the modulino, we have
\begin{align}
 -\ui\ue^{3A+\phi/2}\,&\tr\,\biggl\{\bar{\theta}_{\mathrm{g}}\bigl(
  \ga_{\left(4\right)}\otimes\ga_{m}\bigr)
  \bigl[\vp^{m},\theta_{\mathrm{m}}\bigr]+
  \bar{\theta}_{\mathrm{m}}\bigl(
  \ga_{\left(4\right)}\otimes\ga_{m}\bigr)
  \bigl[\vp^{m},\theta_{\mathrm{g}}\bigr]
  \biggr\}
  \notag\\ &=ab\, \tr\biggl\{\lam\bigl[\vp^{m},\psi^{n}\bigr]\biggr\}
  \Omega_{npq}\,
  \eta_{+}^{\dagger}\bigl\{\ga_{m},\ga^{pq}\bigr\}\eta_{-}
  -ab\, \tr\biggl\{\bar{\lam}\bigl[\vp^{m},\bar{\psi}^{n}\bigr]\biggr\}
  \Omega^{\ast}_{npq}\,
  \eta_{-}^{\dagger}\bigl\{\ga_{m},\ga^{pq}\bigr\}\eta_{+}\notag\\
  &=\frac{\ui\left\lvert\Omega\right\rvert}{2}\omega_{mn}\,
  \tr\biggl\{\lam\bigl[\vp^{m},\psi^{n}\bigr]-\bar{\lam}
  \bigl[\vp^{m},\bar{\psi}^{n}\bigr]\biggr\}.
\end{align}
where we have made use of the cyclicity of the trace
and~\eqref{eq:Omega_contraction}.

Finally, for the modulini Yukawas, we have
\begin{align}
  -\ui\ue^{3A+\phi/2}\,\tr\,\biggl\{\bar{\theta}_{\mathrm{m}}\bigl(
  \ga_{\left(4\right)}\otimes\ga_{m}\bigr)
  \bigl[\vp^{m},\theta_{\mathrm{m}}\bigr]\biggr\}
  =\phantom{-}
  &b^{2}\ue^{\phi/2}\tr\biggl\{\psi^{n}\bigl[\vp^{m},\psi^{r}\bigr]\biggr\}\,
  \Omega_{npq}\Omega_{rst}\,\eta_{+}^{\dagger}\ga^{qp}\ga_{m}
  \ga^{st}\eta_{-}\notag\\
  -&b^{2}\ue^{\phi/2}\tr\biggl\{\bar{\psi}^{n}
  \bigl[\vp^{m},\bar{\psi}^{r}\bigr]\biggr\}\,
  \Omega^{\ast}_{npq}\Omega^{\ast}_{rst}\,\eta_{-}^{\dagger}\ga^{qp}\ga_{m}
  \ga^{st}\eta_{+}.
\end{align}
Making use of~\eqref{eq:Omega_contraction} and the fact that
$\Omega_{mnp}\Omega_{st}^{\phantom{st}p}=0$, this becomes
\begin{multline}
  -\ui\ue^{3A+\phi/2}
  \,\tr\,\biggl\{\bar{\theta}_{\mathrm{m}}\bigl(
  \ga_{\left(4\right)}\otimes\ga_{m}\bigr)
  \bigl[\vp^{m},\theta_{\mathrm{m}}\bigr]\biggr\}
  \\=-\frac{\ue^{\phi/2}}{2}
  \Omega_{mnp}\tr\biggl\{\psi^{m}\bigl[\vp^{n},\psi^{p}\bigr]\biggr\}
  -\frac{\ue^{\phi/2}}{2}
  \Omega^{\ast}_{mnp}\tr\biggl\{\bar{\psi}^{m}\bigl[\vp^{n},
  \bar{\psi}^{p}\bigr]\biggr\}.
\end{multline}

We can now put things together, and the fermionic Lagrangian for a
$\uD 3$ is
\begin{align}
 \cL^{\mathrm{F}}=\tr\biggl\{&-\ui\tilde{K}_{mn}\bar{\psi}^{m}
  \bar{\sig}^{\mu}\partial_{\mu}\psi^{n}-
  \frac{\ui}{g^{2}}\bar{\lambda}\bar{\sig}^{\mu}
  \partial_{\mu}\lambda\notag\\
  &-m_{1/2}\lambda\lambda-m_{1/2}^{\ast}\bar{\lambda}\bar{\lambda}
  -m_{\mathrm{F},m}\lambda\psi^{m}-
  m_{\mathrm{F},m}^{\ast}\bar{\lambda}\bar{\psi}^{m}
  -\frac{1}{2}m_{\mathrm{F},mn}\psi^{m}\psi^{n}
  -\frac{1}{2}m_{\mathrm{F},mn}^{\ast}\bar{\psi}^{m}\bar{\psi}^{n}\notag\\
  &-\ui\,h_{mn}\lambda\psi^{m}\vp^{n}
  -\ui\,h_{mn}^{\ast}\bar{\lambda}\bar{\psi}^{m}\vp^{n}
  -\ui\,h_{mnp}\psi^{m}\psi^{n}\vp^{p}
  -\ui\,h_{mnp}^{\ast}\bar{\psi}^{m}\bar{\psi}^{n}\vp^{p}\biggr\},
  \label{eq:MRVV_action_result}
\end{align}
with
\begin{subequations}
\begin{align}
  \tilde{K}_{mn}=&\frac{\pi}{g_{\us}}\bigl(g_{mn}-\ui\omega_{mn}\bigr),\\
  m_{1/2}=&\frac{\pi}{16 g_{\us}}
  \bigl(\Phi_{+}+\Phi_{-}\bigr)\bigl(\mathrm{Im}\,\tau\bigr)^{1/2}
  G_{-}\cdot\overline{\Omega},\\
  m_{\mathrm{F},m}=&-\frac{\ui\pi}{64 g_{\us}}
  \bigl(\Phi_{+}+\Phi_{-}\bigr)
  \bigl\lvert\Omega\bigr\rvert
  \bigl(g_{ml}+\ui\,\omega_{ml}\bigr)G^{l}_{-nt}\omega^{nt},\\
  m_{\mathrm{F},mn}=&\frac{\pi}{32g_{\us}}\frac{\Phi_{+}+\Phi_{-}}
  {\left(\mathrm{Im}\,\tau\right)^{1/2}}
  \bigl[g_{l\left(m\right.}-\ui\,\omega_{l\left(m\right.}\bigr]
  \Omega_{\left.n\right)pq}G_{-}^{lpq},\\
  h_{mn}=&\frac{2\pi\ui}{g_{\us}}\left\lvert\Omega\right\rvert
  \omega_{mn},\\
  h_{mnp}=&\frac{\pi}{g_{\us}}\frac{1}{\left(\mathrm{Im}\,\tau\right)^{1/2}}
  \Omega_{mnp}.
\end{align}
\end{subequations}

For the case of $\overline{\uD 3}$ branes, we can define the fermionic
degrees of freedom in the same way.  The Lagrangian takes the same
form with the masses modified according to
\begin{subequations}
\begin{align}
  m_{1/2}=&-\frac{\pi}{16 g_{\us}}
  \bigl(\Phi_{+}+\Phi_{-}\bigr)\bigl(\mathrm{Im}\,\tau\bigr)^{1/2}
  \overline{G}_{+}\cdot\overline{\Omega},\\
  m_{\mathrm{F},m}=&\frac{\ui\pi}{64 g_{\us}}
  \bigl(\Phi_{+}+\Phi_{-}\bigr)
  \bigl\lvert\Omega\bigr\rvert
  \bigl(g_{ml}+\ui\,\omega_{ml}\bigr)\overline{G}^{l}_{+nt}\omega^{nt},\\
  m_{\mathrm{F},mn}=&-\frac{\pi}{32g_{\us}}\frac{\Phi_{+}+\Phi_{-}}
  {\left(\mathrm{Im}\,\tau\right)^{1/2}}
  \bigl[g_{l\left(m\right.}-\ui\,\omega_{l\left(m\right.}\bigr]
  {\Omega}_{\left.n\right)pq}\overline{G}_{+}^{lpq}.
\end{align}
\end{subequations}

\section{\label{sec:soft}The soft Lagrangian}

Consider now a general $\cN_{4}=1$ theory.  Such a theory consists of
the supergravity multiplet, vector multiplets giving rise to a gauge
group $G$, and chiral multiplets transforming under various
representations of the gauge group.  The theory is specified by the
K\"ahler function $\cK$, which is a real function of the chiral
superfields, and the superpotential $W$ and gauge kinetic functions
$f$, which are holomorphic in the chiral superfields.  The purpose of
this section is in part to review how the theory for a stack of probe
$\uD 3$s discussed in the previous section can be expressed in terms
of these data in the supersymmetric case.  Additionally, we will argue
that in the non-supersymmetric case, the resulting Lagrangian is
consistent with the spontaneous breaking of supersymmetry.

As discussed in section~\ref{sec:GKP}, an $\cN_{4}=1$ theory is
obtained by taking $G_{\left(3\right)}$ to be $\left(2,1\right)$
primitive (and hence $G_{-}=0$), $\Phi_{-}=0$, the internal metric
$g_{mn}$ to be K\"ahler, and $\tau$ to vary holomorphically over the
internal space.  In this case, all of the masses appearing
in~\eqref{eq:DBI_action_result} and~\eqref{eq:MRVV_action_result}
vanish.  Since our focus is on the interaction of the open strings
with themselves and not the interactions of open strings with
closed-string fluctuations (though such interactions can be
important), we take $\tau$ and the metric to be constant.  After a
constant rescaling of the fields, the low-energy Lagrangian following
from~\eqref{eq:DBI_action_result} and~\eqref{eq:MRVV_action_result}
takes the form
\begin{align}
  \label{eq:4d_D3_Lagrangian}
  \mathcal{L}=\tr\biggl\{&-\frac{1}{4}f_{\mu\nu}f^{\mu\nu}
  -\frac{\vt g^{2}}{32\pi^{2}}
  f_{\mu\nu}\tilde{f}^{\mu\nu}
  -\ui\,\bar{\lambda}\bar{\sig}^{\mu}D_{\mu}
  \lam
  -g_{i\bar{\jmath}}\,D_{\mu}\vp^{i}D^{\mu}\bar{\vp}^{\bar{\jmath}}
  -\ui\,g_{i\bar{\jmath}}\,\bar{\psi}^{\bar{\jmath}}\bar{\sig}^{\mu}D_{\mu}\psi^{i}
  \notag\\
  &-\ui\sqrt{2}g\, g_{i\bar{\jmath}}
  \bigl(\bigl[\vp^{\bar{\jmath}},\psi^{i}\bigr]\lambda
  +\bigl[\vp^{i},\psi^{\bar{\jmath}}\bigr]\bar{\lambda}\bigr)
  +\ui\,g\bigl(\Omega_{ijk}\psi^{i}\psi^{j}\vp^{k}+
  \overline{\Omega}_{\bar{\imath}\bar{\jmath}\bar{k}}
  \bar{\psi}^{\bar{\imath}}\bar{\psi}^{\bar{\jmath}}\bar{\vp}^{\bar{k}}\bigr)\notag\\
  &+\frac{g^{2}}{2}g_{i\bar{\jmath}}g_{k\bar{l}}
  \bigl(\bigl[\vp^{i},\vp^{k}\bigr]
  \bigl[\bar{\vp}^{\bar{\jmath}},\bar{\vp}^{\bar{l}}\bigr]+
  \bigl[\vp^{i},\bar{\vp}^{\bar{l}}\bigr]
  \bigl[\bar{\vp}^{\bar{\jmath}},\vp^{k}\bigr]\bigr)\biggr\},
\end{align}
in which $\tilde{f}_{\left(2\right)}:=\ast_{4}f_{\left(2\right)}$ and
we have made use of the complex structure to separate holomorphic and
anti-holomorphic indices and have used that in our conventions
$\left\lvert\Omega\right\rvert^{2}=8$.  Here the gauge-covariant
derivative is now $D_{\mu}=\partial_{\mu}-\ui
g\left[A_{\mu},\cdot\right]$ due to the field redefinition. 

Let's now compare this to the Lagrangian following from the usual data
of $\cN_{4}=1$ supergravity.  Our interest is in the Lagrangian only
through marginal order and in the rigid supersymmetry limit.  In this case,
the Lagrangian takes the form
\begin{align}
  \label{eq:general_susy_Lagrangian}
  \cL_{\cN_{4}=1}=&-\frac{1}{4}f^{a}_{\mu\nu}f^{a\, \mu\nu} -\frac{\vt
    g^{2}}{32\pi^{2}}f^{a}_{\mu\nu}\tilde{f}^{a\, \mu\nu}
  -\ui\bar{\lambda}^{a}\bar{\sig}D_{\mu}\lambda^{a} -\cK_{I\bar{J}}\,
  D_{\mu}\vp^{I}D^{\mu}\vp^{\bar{J}}
  -\ui\cK_{I\bar{J}}\,\bar{\psi}^{\bar{J}}\bar{\sig}^{\mu}D_{\mu}\psi^{I}\notag\\
  &-\ui\sqrt{2}g\cK_{I\bar{J}}
  \biggl(\bigl(\bar{\vp}^{\bar{I}}T_{\mathrm{r}}^{a}\psi^{J}\bigr)
  \lambda^{a}+\bigl(\bar{\psi}^{\bar{I}}T_{\mathrm{r}}^{a}\vp^{J}\bigr)
  \bar{\lambda}^{a}\biggr) -\frac{g^{2}}{2}
  \bigl(\cK_{I\bar{J}}\bar{\vp}^{\bar{J}}T_{\mathrm{r}}^{a}\vp^{I}\bigr)^{2}
  \notag\\
  &-\cK^{I\bar{J}}W_{I}\bar{W}_{\bar{J}}
  -\frac{1}{2}
  \bigl(W_{IJ}\psi^{I}\psi^{J}+
  \overline{W}_{\bar{I}\bar{J}}\psi^{\bar{I}}\psi^{\bar{J}}\bigr),
\end{align}
in which $W_{I}=\partial_{I}W$, and $W_{IJ}=\partial_{I}\partial_{J}W$
when treating $W$ as a function of the scalar components and
$T_{\mathrm{r}}^{a}$ indicates the generators in the representation
$r$.  Here $\cK_{I\bar{J}}$ is assumed to be non-singular at
$\vp^{I}=0$ and is evaluated at that point.

Comparing to~\eqref{eq:4d_D3_Lagrangian}, we immediately make the
well-known identification of the matter-field metric with the internal
metric $\cK_{I\bar{J}}\to g_{i\bar{\jmath}}$.  To deduce the
superpotential that corresponds to~\eqref{eq:4d_D3_Lagrangian}, we
note that we can write
\begin{equation}
  \ui g\,\Omega_{ijk}\tr\bigl\{\psi^{i}\psi^{j}\vp^{k}\bigr\}
  =\frac{\ui g}{2}\Omega_{ijk}
  \tr\bigl\{\psi^{i}\bigl[\psi^{j},\vp^{k}\bigr]\bigr\}
  =-\frac{1}{2}f^{abc}\Omega_{ijk}\psi^{i}_{a}\psi^{j}_{b}\vp^{k}_{c},
\end{equation}
where we have normalized the generators according to
$\tr\left(T^{a}T^{b}\right)=\delta^{ab}$ and defined the structure
constants $\left[T^{a},T^{b}\right]=\ui f^{abc}T^{c}$.  Thus the
Yukawa couplings not involving the gaugino follow from
\begin{equation}
  \label{eq:N4_superpotential}
  W_{\cN_{4}=4}=\frac{g}{3!}f^{abc}\Omega_{ijk}\vp^{i}_{a}\vp^{i}_{b}\vp^{k}_{c}
  =-\frac{\ui g}{3}\Omega_{ijk}\tr\bigl\{\vp^{i}\vp^{j}\vp^{k}\bigr\},
\end{equation}
which is the usual superpotential used to describe $\cN_{4}=4$ super
Yang-Mills theory in $\cN_{4}=1$ language.  From this superpotential,
we find the $F$-term potential
\begin{equation}
  V_{F}=g^{i\bar{\jmath}}W_{i}\bar{W}_{\bar{\jmath}}=-g^{2}g_{i\bar{\jmath}}
  g_{k\bar{l}}\tr\bigl\{\bigl[\vp^{i},\vp^{k}\bigr]
  \bigl[\bar{\vp}^{\bar{\jmath}},\bar{\vp}^{\bar{l}}\bigr]\bigr\}.
\end{equation}
Adding this to the $D$-term potential
\begin{equation}
  V_{D}=-\frac{g^{2}}{2}
  \bigl(g_{i\bar{\jmath}}\bigl[\vp^{i},\bar{\vp}^{\jmath}\bigr]\bigr)^{2},
\end{equation}
and making use of the Jacobi identity we recover the scalar potential
appearing in~\eqref{eq:4d_D3_Lagrangian}.

We now turn to the more general case in which the geometry no longer
satisfies the conditions for supersymmetry.  An important distinction
between the supersymmetric and non-supersymmetric cases, as discussed
in section~\ref{sec:GKP}, is that once non-supersymmetric fluxes are
introduced to the geometry, the equations of motion imply that the
internal metric $g_{mn}$ will generically no longer be Hermitian with
respect to the unperturbed complex structure.  Indeed, there is no
guarantee at this level that the internal metric is even either
complex or symplectic.  However, we can make use of the almost complex
structure and pre-symplectic structures that are defined, at least
locally, by the Killing spinor of the non-perturbed
geometry~\eqref{eq:structure}.  Note that although the same spinor is
used, it will not generically satisfy the Killing spinor equations of
the perturbed geometry.  Furthermore, since the internal gamma
matrices are defined in terms of the vielbein
\begin{equation}
  \ga_{m}=e_{m}^{\phantom{m}\ul{n}}\ga_{\ul{n}},
\end{equation}
and these vielbein are perturbed according to the perturbation of the
metric, the almost complex structure and pre-symplectic structure are
not equal to their non-perturbed counterparts.  In what follows, we
will make use of this almost complex structure to locally define
holomorphic and anti-holomorphic indices, keeping in mind that the
structure is not expected to be integrable.

Before discussing the Lagrangian resulting from the stack of $\uD 3$s,
let us review the impact that the breaking of supersymmetry can have
on the system.  As briefly mentioned in the introduction, there are
two ways that supersymmetry can be broken in a theory.  The first is
explicit breaking in which the theory is altered by changing the
action (which may be accompanied by changing the field content) such
that it no longer respects any supersymmetry transformations.  The
second way is by spontaneous or dynamical supersymmetry breaking in
which the theory is invariant under supersymmetry transformations, but
the supercharges do not annihilate the state being considered,
typically a meta-stable false vacuum.  This latter case is, from a
phenomenological standpoint, more interesting since spontaneous
breaking restricts the sorts of terms that can appear in the resulting
effective field theory so that certain operators, such as scalar
masses, are protected from large quantum corrections.  In the case of
spontaneous breaking, the effective field theory may not have manifest
supersymmetry, but instead supersymmetry may be realized only
non-linearly.  This is similar to the case of the spontaneous breaking
of bosonic symmetries.  The effective low-energy Lagrangian in such a
case is a non-linear $\Sigma$-model in which the symmetry, though
spontaneously broken and realized only non-linearly, greatly restricts
low-energy physics.

\begin{figure}[t]
  \begin{center}
    \begin{fmffile}{tadpole}
      \begin{fmfgraph*}(100,100)
        \fmfleft{i}
        \fmfright{r}
        \fmf{vanilla,label=$\vp^{i}$}{i,v1}
        \fmf{phantom,tension=2}{v2,r}
        \fmf{vanilla,left,label=$\vp^{j}$,tension=.5}{v1,v2}
        \fmf{vanilla,left,label=$\bar{\vp}^{\bar{k}}$,tension=.5}{v2,v1}
        \fmfdot{v1,v2}
      \end{fmfgraph*}
    \end{fmffile}
    \caption{\label{fig:tadpole}A Feynman diagram demonstrating hard
      breaking of supersymmetry.  Integration over the internal
      momentum gives rise to quadratic dependence on the UV regulator.
      However, if $\vp^{i}$ is not a gauge singlet, then gauge
      invariance forces this diagram to vanish.  Note that the
      holomorphic $a$-terms do not give rise to such hard breaking
      since kinetic terms, schematically represented by the insertion
      in the loop, do not mix holomorphic with holomorphic fields.}
  \end{center}
\end{figure}
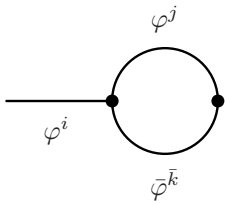

In a typical model of supersymmetry
breaking\footnote{See~\cite{Martin:1997ns,*Chung:2003fi} for reviews
  and~\cite{Soni:1983rm,*Nilles:1983ge,*Brignole:1993dj,*Brignole:1995fb,*Brignole:1997dp}
  for early treatments of gravity mediation which is the mechanism
  most relevant for the discussion here.} supersymmetry is broken
spontaneously in a particular sector of a theory by a non-vanishing
expectation value for an $F$-term or a $D$-term (other possibilities
exist if one is willing to give up Lorentz invariance).  In order to
avoid a phenomenologically unacceptable spectrum, supersymmetry
breaking is typically assumed to occur in a ``hidden'' sector, rather
than in the visible sector of interest.  The effects of the breaking
are then mediated by a (not necessarily distinct) sector known as the
messenger sector.  Upon integrating out the hidden and messenger
sectors, the resulting visible sector does not possess manifest
supersymmetry.  However, the Lagrangian is non-generic in that only
the relevant operators do not obey supersymmetry relations.  That is,
after breaking supersymmetry, the visible sector Lagrangian in the
rigid limit takes the form
\begin{equation}
  \cL_{\mathrm{vis}}=\cL_{\mathrm{susy}}+\cL_{\mathrm{soft}},
\end{equation}
where $\cL_{\mathrm{susy}}$ linearly preserves supersymmetry, while
$\cL_{\mathrm{soft}}$ does not, but has no operators of dimension
greater than three.  In general, $\cL_{\mathrm{soft}}$ takes the
schematic form
\begin{equation}
  \label{eq:soft_Lagrangian}
  \cL_{\mathrm{soft}}\sim t_{i}\vp^{i}+b_{ij}\vp^{i}\vp^{j}
  +m^{2}_{i\bar{\jmath}}\vp^{i}\bar{\vp}^{\bar{\jmath}}+
  m_{1/2}\lambda\lambda
  +m_{i}\lambda\psi^{i}+a_{ijk}\vp^{i}\vp^{k}\vp^{k}
  +c_{ij\bar{k}}\vp^{i}\vp^{j}\bar{\vp}^{\bar{k}}+\mathrm{h.c.}
\end{equation}
Other operators can be shuffled into a redefinition of the
superpotential as we will do below.  We also note the existence of the
potentially unfamiliar term $m_{i}\lambda\psi^{i}$ that can be present
for adjoint-valued fields, while in more phenomenologically viable
constructions, such fields typically do not exist and so this term is
absent.  The operators are denoted ``soft'' since although the absence
of supersymmetry implies less protection against quantum corrections,
most of the operators in $\mathcal{L}_{\mathrm{soft}}$ only depend
logarithmically on the scale of ultraviolet physics.  However, some of
these operators may break supersymmetry in a hard manner.  In
particular, the operator
$c_{ij\bar{k}}\vp^{i}\vp^{j}\bar{\vp}^{\bar{k}}$ will produce
quadratically divergent tadpole graphs such as that appearing in
figure~\ref{fig:tadpole}.  Fortunately, gauge invariance implies that
such a graph can only be non-vanishing if one of the fields is a
singlet and it is easy to argue, as we do below, that these couplings
are absent for the singlets in the theory.

The structure of the Lagrangian in the case of spontaneous breaking is
to be contrasted with the generic Lagrangian in the case of explicit
breaking.  In the latter case, one would expect from a Wilsonian
standpoint that the resulting Lagrangian would have no special
structure and instead consist of all scalar operators consistent with
gauge invariance.  In particular, generic explicit breaking should
also lead to non-supersymmetric marginal deformations of the
Lagrangian.

We now return to the case of $\uD 3$s probing a flux compactification.
As can be seen from the results of the previous section, a general
perturbation to a supersymmetric flux compactification alters the
Lagrangian~\eqref{eq:general_susy_Lagrangian} in a more drastic way
than the simple addition of~\eqref{eq:soft_Lagrangian}.  It
particular, a general perturbation from $\cN_{4}=1$ GKP will modify
the marginal couplings, namely the kinetic terms, Yukawa couplings and
$\vp^{4}$.  However, by making use of the perturbed
structures~\eqref{eq:structure}, it follows immediately that the
marginal operators take the same form as they do
in~\eqref{eq:4d_D3_Lagrangian} when written in terms of these
structures.  As stated previously, in general the structures are not
expected to be integrable and so $g_{i\bar{\jmath}}$ is not expected
to be K\"ahler.  Insofar as we are interested only in relevant and
marginal operators, this will not affect the Lagrangian and so still
the marginal operators follow from the
superpotential~\eqref{eq:N4_superpotential}.  We return to this point
of non-K\"ahlerity in section~\ref{sec:conclusion}.

In terms of these renormalized fields and the local almost complex
structure, we can rewrite the Lagrangians~\eqref{eq:DBI_action_result}
and~\eqref{eq:MRVV_action_result} as
\begin{align}
  \label{eq:final_D3_result}
  \mathcal{L}=
  \tr\biggl\{&-\frac{1}{4}f_{\mu\nu}f^{\mu\nu}
  -\frac{\vt g^{2}}{32\pi^{2}}f_{\mu\nu}\tilde{f}^{\mu\nu}
  -\ui\bar{\lambda}^{a}\bar{\sig}D_{\mu}\lambda^{a}
  -g_{i\bar{\jmath}}\,
  D_{\mu}\vp^{i}D^{\mu}\bar{\vp}^{\bar{\jmath}}
  +\ui\,g_{i\bar{\jmath}}\,\bar{\psi}^{\bar{\jmath}}
  \bar{\sig}^{\mu}D_{\mu}\psi^{i}\notag\\
  &-\ui\sqrt{2}g\, g_{i\bar{\jmath}}
  \bigl(\bigl[\vp^{\bar{\jmath}},\psi^{i}\bigr]\lambda
  +\bigl[\vp^{i},\psi^{\bar{\jmath}}\bigr]\bar{\lambda}\bigr)
  +\ui\,g\bigl(\Omega_{ijk}\psi^{i}\psi^{j}\vp^{k}+
  \overline{\Omega}_{\bar{\imath}\bar{\jmath}\bar{k}}
  \bar{\psi}^{\bar{\imath}}\bar{\psi}^{\bar{\jmath}}\bar{\vp}^{\bar{k}}\bigr)\notag\\
  &+\frac{g^{2}}{2}g_{i\bar{\jmath}}g_{k\bar{l}}
  \bigl(\bigl[\vp^{i},\vp^{k}\bigr]
  \bigl[\bar{\vp}^{\bar{\jmath}},\bar{\vp}^{\bar{l}}\bigr]+
  \bigl[\vp^{i},\bar{\vp}^{\bar{l}}\bigr]
  \bigl[\bar{\vp}^{\bar{\jmath}},\vp^{k}\bigr]\bigr)\notag\\
  &-\bigl(t_{i}\vp^{i}+t^{\ast}_{\bar{\imath}}\bar{\vp}^{\bar{\imath}}\bigr)
  -\frac{1}{2}\bigl(b_{ij}\vp^{i}\vp^{j}+
  b^{\ast}_{\bar{\imath}\bar{\jmath}}
  \bar{\vp}^{\bar{\imath}}\bar{\vp}^{\bar{\jmath}}\bigr)
  -m_{i\bar{\jmath}}^{2}\vp^{i}\vp^{\bar{\jmath}}\notag\\
  &-\frac{\ui}{3!}\bigl(
  a_{ijk}\vp^{i}\vp^{j}\vp^{k}+
  a^{\ast}_{\bar{\imath}\bar{\jmath}\bar{k}}
  \bar{\vp}^{\bar{\imath}}\bar{\vp}^{\bar{\jmath}}\bar{\vp}^{\bar{k}}\bigr)
  -\frac{\ui}{2!}\bigl(
  c_{ij\bar{k}}\vp^{i}\vp^{j}\bar{\vp}^{\bar{k}}+
  c^{\ast}_{\bar{\imath}\bar{\jmath}{k}}
  \bar{\vp}^{\bar{\imath}}\bar{\vp}^{\bar{\jmath}}{\vp}^{{k}}\bigr)\notag\\
  &-\bigl(m_{1/2}\lambda\lambda+m_{1/2}^{\ast}\bar{\lambda}\bar{\lambda}\bigr)
  -\bigl(m_{i}\lambda\psi^{i}
  +m^{\ast}_{\bar{i}}\bar{\lambda}\bar{\psi}^{\bar{i}}\bigr)
  -\frac{1}{2}\bigl(\mu_{ij}\psi^{i}\psi^{j}
  +\mu^{\ast}_{\bar{\imath}\bar{\jmath}}\bar{\psi}^{\bar{\imath}}
  \bar{\psi}^{\bar{\jmath}}\bigr)
  \biggr\},
\end{align}
in which for the $\uD 3$ case (recall
$g^{-2}=\frac{2\pi}{g_{\us}}\mathrm{Im}\, \tau$)
\begin{subequations}
\begin{align}
  t_{i}=&\sqrt{\frac{2\pi}{g_{\us}}}\frac{1}{\ell_{\us}^{2}}\partial_{i}\Phi_{-},
  \\
  b_{ij}=&\partial_{i}\partial_{j}\Phi_{-},\\
  m_{i\bar{\jmath}}^{2}=&\partial_{i}\bar{\partial}_{\bar{\jmath}}\Phi_{-},\\
  a_{ijk}=&-\frac{g^{2}}{2}\sqrt{\frac{2\pi}{g_{\us}}}
  \bigl(G_{-}+\overline{G}_{-}\bigr)_{ijk},\\
  c_{ij\bar{k}}=&-\frac{g^{2}}{2}\sqrt{\frac{2 \pi}{g_{\us}}}
  \bigl(G_{-}+\overline{G}_{-}\bigr)_{ij\bar{k}},\\
  m_{1/2}=&g\sqrt{\frac{2\pi}{g_{\us}}}\frac{\Phi_{+}+\Phi_{-}}{32}
  G_{-}\cdot\overline{\Omega},\\
  m_{i}=&\frac{g}{8\sqrt{2}}\sqrt{\frac{2\pi}{g_{\us}}}
  \bigl(\Phi_{+}+\Phi_{-}\bigr)G_{-\,ij}^{\phantom{-\,ij}j},\\
  \mu_{ij}=&g\sqrt{\frac{2\pi}{g_{\us}}}
  \frac{\Phi_{+}+\Phi_{-}}{32}
  G^{\phantom{-\left(i\right.}kl}_{-\left(i\right.}
  \Omega^{\phantom{\left.i\right)}}_{\left.j\right)kl}.
\end{align}
\end{subequations}
As mentioned above, the general Lagrangian contains a holomorphic mass
term for the fermions.  Such a term can be absorbed into a superpotential as discussed in~\cite{Camara:2003ku},
\begin{equation}
  \label{eq:new_superpotential}
  W=W_{\mathrm{\cN}_{4}=4}+\frac{1}{2}\mu_{ij}\vp^{i}\vp^{j},
\end{equation}
in which $W_{\cN_{4}=4}$ is given by~\eqref{eq:N4_superpotential}.

The gauge group on the $\uD 3$ branes is a semi-direct product
$\U{N}=\SU{N}\rtimes\U{1}$.  Since the adjoint of $\U{1}$ is trivial,
this implies the existence of gauge singlets on the worldvolume (for
example, the center-of-mass of the $\uD 3$ branes in the internal
space) and hence we must check if the terms like $c_{ij\bar{k}}$ lead
to the hard breaking of supersymmetry.  To this end, we expand the
open-string fields in terms of the generators of the gauge group,
e.g. $\vp^{i}=\vp^{i}_{a}T^{a}$, and denote the $\U{1}$ generator by
$T^{0}$.  $c_{ij\bar{k}}$ is anti-symmetric in $i$ and $j$, so the
terms $\vp^{i}_{0}$ or $\vp^{j}_{0}$ automatically vanish as $T^{0}$
commutes with everything.  The term involving $\vp^{\bar{k}}_{0}$ also
vanishes since we have, for $a,b\neq 0$,
\begin{equation}
  c_{ij\bar{k}}\vp_{a}^{i}\vp^{j}_{b}\bar{\vp}^{\bar{k}}_{0}\tr\bigl\{
  T_{a}T_{b}T_{0}\bigr\}
  \sim
  c_{ij\bar{k}}\vp_{a}^{i}\vp^{j}_{b}\bar{\vp}^{\bar{k}}_{0}
  f^{abc}\tr\bigl\{T^{c}T^{0}\bigr\}=0,
\end{equation}
which follows from $T^{c}\neq T^{0}$.  Therefore the couplings of the
type $c_{ij\bar{k}}$ vanish when they involve singlets and so do not
introduce any hard breaking.

Let us pause to emphasize that many of these operators have appeared
elsewhere in the literature.  The relevant operators as well as the
$\vp^{4}$ operator appeared in the weak-warping limit
in~\cite{Camara:2003ku}.  These operators also appeared with more
general warping in the Abelian case in~\cite{Grana:2003ek} as did a
subset of them in~\cite{Burgess:2006mn}.  Finally, some of these
operators can be deduced via worldsheet
methods~\cite{Lust:2004fi,*Lust:2004dn}. However, in these works, the
expression of such terms in terms of softly-broken $\cN_{4}=1$
language made use of the existence of an underlying complex structure
(i.e. when the underlying metric is K\"ahler) while here we have
expressed the Lagrangian in terms of a softly broken supersymmetric
Lagrangian (including Yukawa couplings that were not considered in
some previous work) in more general cases.

The superpotential~\eqref{eq:new_superpotential} is of course
holomorphic, but it is holomorphic with respect to a perturbed complex
structure.  Said differently,~\eqref{eq:new_superpotential} is not
holomorphic in the fields of the $\uD 3$ probing the non-perturbed GKP
compactification but instead holomorphic in fields after a
non-holomorphic field redefinition.  This implies that although the
Lagrangian~\eqref{eq:final_D3_result} describes a theory of a
spontaneously broken rigid supersymmetry, this supersymmetry is not
the same as the supersymmetry preserved by GKP. Instead the
supercharges that are treated as spontaneously broken
in~\eqref{eq:final_D3_result} is some linear combination of the
supercharges preserved by GKP and those that are not,
\begin{equation}
  Q_{\alpha}\sim Q_{\alpha}^{\mathrm{GKP}}+\sum c_{a} Q^{a}_{\alpha},
\end{equation}
where the coefficients $c_{a}$ are of the same order as the
perturbation to GKP and we have kept the spinor index explicit.  We
note that a similar phenomenon must occur even with certain
supersymmetric perturbations. Changes to the complex structure (which
of course must involve either complex structure moduli that are not
fixed by fluxes or a modification of the fluxes as well) implies that
a different spinor $\eta_{+}$ is annihilated by the new
anti-holomorphic $\ga$-matrices and so correspondingly the preserved
supercharges is shifted.

Another way to see the shift in supercharges is in terms of the
gravitini.  Type-IIB is a theory with 32 supercharges, and so a
toroidal compactification to 4d gives eight gravitini
$\psi_{\mu\alpha}^{I}$ where $I=1,\cdots, 8$.  On a Calabi-Yau with
strictly $\SU{3}$ holonomy, only two of these gravitini remain light,
and the other six can be thought of as being lifted to the
Kaluza-Klein.  Once supersymmetric fluxes have been added, the
remaining gravitino is (at least in generic cases) lifted by that
flux, leaving a single light gravitino
$\psi_{\mu\alpha}^{\mathrm{GKP}}$ (without fluxes, the geometry cannot
distinguish between $\uD 3$-branes and $\overline{\uD 3}$-branes, so
the gravitino lifted by the fluxes must correspond to the supercharges
preserved by an $\overline{\uD 3}$-brane).  Finally, once
non-supersymmetric fluxes have been added, the remaining gravitino
will also be lifted (see, for example~\cite{DeWolfe:2002nn}).
Schematically, and neglecting warping effects, the mass is similar to
that for the gaugino discussed above
\begin{equation}
  \label{eq:gravitino_mass}
  m_{3/2}\sim \int\eta_{+}^{\dagger}\overline{\slashed{G}}\eta_{-},
\end{equation}
which follows from the 10d gravitino action and depends on the
$\left(0,3\right)$ ISD flux.  However, generically all Hodge types of
fluxes are sourced and this will give rise to mixing between
$\psi_{\mu\alpha}^{\mathrm{GKP}}$ and the gravitini lifted by the
geometry and flux.  Due to this mixing, the lightest gravitino will
not be the GKP gravitino, but instead will include an admixture of
the gravitini lifted by GKP itself.  Although we leave a more precise
treatment (for example, the incorporation of the compactification
effects required to ensure a finite Kaluza-Klein scale and
that~\eqref{eq:gravitino_mass} is well-defined) for future work, this
mixture of gravitini is another way of understanding the physics of
why~\eqref{eq:new_superpotential} is not holomorphic in the
unperturbed fields.

If supersymmetry is broken by the addition of $\overline{\uD
  3}$-branes, then the supersymmetric state obtained by the system
after the decay of such branes may not be the same as supersymmetric
state to which the anti-branes were originally added.  For example, in
the KS system the $\overline{\uD 3}$s decay, via $\mathrm{NS5}$s, into
flux and $\uD 3$-branes that were not present in the original KS
geometry~\cite{Kachru:2002gs}.  This system has, due to the change in
flux, a different complex structure and hence a different
supersymmetry than the one preserved by the geometry before the
addition of the anti-branes.  Generically, one would again expect that
the lightest gravitino is not quite the gravitino gauging the
supersymmetry in this final state, but it would be worthwhile to
understand this in detail\footnote{We thank T. Wrase for discussions
  related to this point.}.

We close this section by noting that the softness of the $\uD
3$-action is independent of the background that the $\uD 3$-branes are
probing.  In the approximation scheme of our analysis, the marginal
operators are controlled exclusively by the internal metric.  Although
in the above analysis we considered small perturbations away from GKP,
we can always perform a local field redefinition so that the
matter-field metric always takes a form proportional to
$\delta_{i\bar{\jmath}}$.  This field redefinition will also cause the
Yukawa-couplings and $\vp^{4}$ potentials to take the form that they
do in~\eqref{eq:general_susy_Lagrangian}.

\section{\label{sec:goldstino}An anti-brane goldstino}

The result of the previous section is that through marginal order, a
stack of $\uD 3$-branes probing a perturbation of an $\cN_{4}=1$ GKP
compactification in the supergravity limit experiences the breaking of
supersymmetry softly.  Although soft breaking and spontaneous breaking
are not equivalent (indeed, as discussed previously non-zero
$c_{ij\bar{k}}$ which may be present will introduce hard breaking in
certain other models), soft breaking is a very non-generic feature of
models of explicit breaking.  We thus take the softness of the $\uD 3$
action as evidence that the non-supersymmetric flux, and therefore
what is giving rise to that flux, may break supersymmetry
spontaneously (although other explanations may be possible).  If
supersymmetry is indeed spontaneously broken, then there must exist a
fermion that is massless in the $m_{\mathrm{p}}\to\infty$ limit.  In
this section, we consider the case where the fluxes result as a
backreaction of an $\overline{\uD 3}$ and argue for the presence of
such a goldstino in the spectrum of $\overline{\uD 3}$-fluctuations.

To this end, we consider the effective
Lagrangians~\eqref{eq:DBI_action_result}
and~\eqref{eq:MRVV_action_result} for the case of an $\overline{\uD
  3}$-brane probing a GKP compactification that exhibits $\cN_{4}=1$
before the addition of the $\overline{\uD 3}$.  In that case, the
Lagrangian again takes the form~\eqref{eq:final_D3_result} with
\begin{subequations}
\label{eq:anti_d3_terms}
\begin{align}
  g^{-2}=&\frac{2\pi}{g_{\us}}\mathrm{Im}\,\tau,\\
  \vt=&-\frac{16\pi^{3}}{g_{\us}}\mathrm{Re}\,\tau,\\
  t_{i}=&\sqrt{\frac{2\pi}{g_{\us}}}\frac{1}{\ell_{\us}^{2}}\partial_{i}\Phi_{+},
  \\
  b_{ij}=&\partial_{i}\partial_{j}\Phi_{+},\\
  m_{i\bar{\jmath}}^{2}=&\partial_{i}\bar{\partial}_{\bar{\jmath}}\Phi_{+},\\
  a_{ijk}=&-\frac{g^{2}}{2}\sqrt{\frac{2\pi}{g_{\us}}}
  \bigl(G_{+}+\overline{G}_{+}\bigr)_{ijk},\\
  c_{ij\bar{k}}=&-\frac{g^{2}}{2}\sqrt{\frac{2 \pi}{g_{\us}}}
  \bigl(G_{+}+\overline{G}_{+}\bigr)_{ij\bar{k}},\\
  m_{1/2}=&-g\sqrt{\frac{2\pi}{g_{\us}}}\frac{\Phi_{+}+\Phi_{-}}{32}
  \overline{G}_{+}\cdot\overline{\Omega},\\
\intertext{\newpage}
  m_{i}=&-\frac{g}{8\sqrt{2}}\sqrt{\frac{2\pi}{g_{\us}}}
  \bigl(\Phi_{+}+\Phi_{-}\bigr)\overline{G}_{+\,ij}^{\phantom{-\,ij}j},\\
  \mu_{ij}=&-g\sqrt{\frac{2\pi}{g_{\us}}}
  \frac{\Phi_{+}+\Phi_{-}}{32}
  \overline{G}^{\phantom{+\left(i\right.}kl}_{+\left(i\right.}
  \Omega^{\phantom{\left.i\right)}}_{\left.j\right)kl}.
\end{align}
\end{subequations}
In what follows, we will for simplicity consider a single
$\overline{\uD 3}$-brane so that $a$ and $c$ both vanish\footnote{In
  the case of multiple $\overline{\uD 3}$s, the goldstino is most
  likely related to the $\U{1}$ part of the fermionic mode that we
  identify below, just as for multiple D-branes the goldstone is the
  center-of-mass.}.  Here $g_{i\bar{\jmath}}$ is the K\"ahler metric
of the unperturbed geometry and $\Omega_{ijk}$ is the form associated
with the complex structure.  Now, in addition to an $\cN_{4}=1$ GKP
compactification having $G_{-}=0$, the non-vanishing $G_{+}$ part is
restricted to be $\left(2,1\right)$ and primitive and as a
consequence,
\begin{equation}
  m_{1/2}=0,\qquad m_{i}=0.
\end{equation}
That is, the gaugino on the $\overline{\uD 3}$ is massless.  Note that
it was important that both $m_{1/2}$ and $m_{i}$ vanished; even if
$m_{1/2}$ vanished but $m_{i}$ were non-vanishing then upon
diagonalization of the mass matrix, there would generically not be any
massless mode.

The massless fermions on a ${\uD p}$-brane in flat space can be
considered as goldstini associated with the spontaneous breaking of 16
supercharges.  However, just as the action for a goldstone boson is
restricted, the action for a goldstino $\chi$ in $R^{3,1}$ takes the
Akulov-Volkov form~\cite{Volkov:1972jx, Volkov:1973ix},
\begin{equation}
  \label{eq:AV_Lagrangian}
  S_{\mathrm{AV}}=-\frac{f^{2}}{2}\int\ud^{4}x\,
  \det\biggl[\delta^{\mu}_{\nu}+\frac{\ui}{f^{2}}
  \bigl(\bar{\chi}\bar{\sig}^{\mu}\partial_{\nu}\chi
  +\chi\sig^{\mu}\partial_{\nu}\chi\bigr)\biggr],
\end{equation}
in which $f$ is related to the scale of the breaking of supersymmetry.
Although this action does not contain a full multiplet, it is
invariant under the transformation
\begin{equation}
  \delta_{\ep} \chi=f\ep-\frac{\ui}{f}\bigl(\chi\sig^{\mu}\bar{\ep}
  -\ep\sig^{\mu}\bar{\chi}\bigr)\partial_{\mu}\chi,
\end{equation}
in which $\ep$ is an arbitrary constant spinor.  Moreover, this
transformation reproduces the usual supersymmetry algebra and
so~\eqref{eq:AV_Lagrangian} realizes supersymmetry non-linearly.

If the fermionic modes on a $\uD p$ are to be interpreted as
goldstini, then their action must be similarly constrained.  The
action for a single $\uD p$-brane in flat space can be expanded out to
higher order in fermions.  In flat space, the $\ka$-fixed action takes
the form~\cite{Aganagic:1996nn} (matching to our conventions)
\begin{align}
  \label{eq:non_linear_Dp_action}
  S_{\uD p}=&-\tau_{\uD p}\int\ud^{p+1}
  \xi\sqrt{-\det\left({M}_{\alpha\beta}\right)},
  \\
  M_{\alpha\beta}=&{\eta}_{\alpha\beta}+\ell_{\us}^{2}f_{\alpha\beta}
  +\ell_{\us}^{4}\partial_{\alpha}\vp^{i}\partial_{\beta}\vp^{i}
  -\ui\ell_{\us}^{4}\bar{\theta}\bigl({\Ga}_{\alpha}
  +\ell_{\us}^{2}{\Ga}_{i}\partial_{\alpha}\vp^{i}\bigr)\partial_{\beta}\theta
  -\frac{1}{4}\ell_{\us}^{8}
  \bigl(\bar{\theta}{\Ga}^{M}\partial_{\alpha}\theta\bigr)
  \bigl(\bar{\theta}{\Ga}_{M}\partial_{\beta}\theta\bigr).\notag
\end{align}
The term that is quadratic order in fermions is the
action~\eqref{eq:abelian_Martucci}.  In addition to the linearized
supersymmetry transformations corresponding to the supercharges that
the $\uD p$ preserves, it also realizes another set of supersymmetries
non-linearly, as detailed in~\cite{Aganagic:1996nn}.
Unlike~\eqref{eq:AV_Lagrangian}, the action for a $\uD p$ brane
realizes some supersymmetry in a linear way and therefore we should
not expect to recover precisely~\eqref{eq:AV_Lagrangian} and
indeed,~\eqref{eq:non_linear_Dp_action} is closed under the
non-linearly realized supersymmetry only once the bosonic terms are
included~\cite{Aganagic:1996nn}. Nevertheless, the corresponding
supersymmetry transformations realize the supersymmetry algebra and
$\theta$, which appears non-linearly in the supersymmetry
transformations, ought to be identified with the goldstini associated
with the spontaneous breaking of 16 supercharges by the $\uD p$-brane.

When moving to flux backgrounds, the extension of the action to
higher-order in fermions becomes more complicated (see
e.g.~\cite{Simon:2011rw} for a review of related issues).  However,
the physics of the situation remains the same: D-branes spontaneously
break supersymmetry and the worldvolume fermions are the corresponding
goldstini.  We are not aware of a presentation of the higher-order
fermionic action in a flux background that is as readily applicable
as~\eqref{eq:abelian_Martucci} or~\eqref{eq:non_linear_Dp_action}, but
from the higher-order terms presented in~\cite{Aganagic:1996nn} and
the expansion of the Ramond-Ramond superfields as presented in, for
example,~\cite{Marolf:2003ye,*Marolf:2003vf}, it is clear that it must
involve products of bilinears of the type that appear
in~\eqref{eq:abelian_Martucci}, at least in the absence of scalar
fluctuations.  When $\theta$ is pure gaugingo then the Hodge-types of
background fluxes and the property that
$\bar{\theta}\hat{\Ga}_{M_{1}\cdots M_{p}}\theta$ vanishes if $p\neq
0,3,7,10$ imply that the only non-vanishing bilinears are the derivative
terms.  Therefore, the higher-order fermionic action is expected to
take the form~\eqref{eq:non_linear_Dp_action} when $\theta$ is pure
gaugino, with some modifications due to warping.  By comparing the
scale of the constant term in the action to that of the kinetic term
of the gaugino, we find
\begin{equation}
  f^{2}=\tau_{\uD 3}\ue^{4A}\sim\ell_{\us}^{-4}\ue^{4A},
\end{equation}
which is the familiar statement that the scale of supersymmetry is
warped down from the string scale~\cite{Giddings:2001yu,Kachru:2003aw}.

Finally, let's consider the case in which, instead of probing an
$\cN_{4}=1$ GKP compactification, the $\overline{\uD 3}$ probes a
compactification with non-vanishing $\left(0,3\right)$ or primitive
$\left(1,2\right)$ flux.  Although such flux is still ISD and so the
internal space is K\"ahler, supersymmetry is no longer preserved by
the background, and so the $\overline{\uD 3}$ is not, by itself,
responsible for the breaking of the supersymmetry preserved by the
primitive $\left(2,1\right)$ flux. From~\eqref{eq:anti_d3_terms} it is
clear that in this case the gaugino will no longer be massless, which
is consistent with the fact that the goldstino cannot be exclusively
an $\overline{\uD 3}$ mode and consistent also with the interpretation
of the $\overline{\uD 3}$ gaugino as the goldstino when probing a
supersymmetric compactification.  A possible objection to this line of
reasoning comes from considering the limit in which all of the flux
vanishes.  Then according to~\eqref{eq:anti_d3_terms}, all of the
fermions on the $\overline{\uD 3}$ are massless, yet the goldstino
cannot be purely an open-string mode since the geometry itself breaks
three of the supercharges preserved by the $\overline{\uD 3}$ and so
one should expect some closed-string component to the goldstino (note
that the same issue arises for $\uD 3$-branes as the geometry itself
cannot distinguish between the charges).  However, one can imagine
going to a region in moduli space where the internal volume is very
large and flat and so the $\overline{\uD 3}$ is, to good
approximation, probing flat 10d space, in which case the
interpretation of the $\overline{\uD 3}$ modulini modes as goldstini
is appropriate.  It is therefore not surprising that the modulini will
be massless in other regions of moduli space.  Presumably, the
goldstino is at all points in moduli space a mixture of open- and
closed-string modes.  It would be interesting to confirm this fact by
understanding the super-Higgs mechanism in such cases.  Note that this
is again entirely analogous to what occurs in the bosonic sector: the
Calabi-Yau itself generically has no isometries and yet there are
still massless bosons that are neatly associated with goldstones modes
associated with the spontaneous breaking of translational symmetry in
the large-radius limit.

To summarize this section, we have identified the $\overline{\uD 3}$
gaugino as a candidate for the goldstino associated with the
spontaneous breaking of supersymmetry.  Although more work is required
to rigorously demonstrate this, the gaugino is massless when an
$\overline{\uD 3}$ probes $\cN_{4}\ge 1$ GKP compactifications and
massive when probing $\cN_{4}=0$ GKP compactifications, as is expected
from such a goldstino.

\newpage

\section{\label{sec:conclusion}Discussion and concluding remarks}

In this work, we have presented some circumstantial evidence that
$\overline{\uD 3}$s spontaneously break supersymmetry in a flux
compactification, contrary to some common folklore which claims that
they are an explicit source of breaking.  Although this evidence is
not conclusive, it approaches a coherent story about the breaking of
supersymmetry by anti-branes.  In this final section, we summarize
these arguments, discuss some possible objections, and lay out some
directions for future work.  Although many of the arguments here refer
explicitly to $\overline{\uD 3}$-branes, they apply to other sources
of supersymmetry breaking as well.  However, since we have been able
to identify a candidate goldstino in the case of $\overline{\uD 3}$s,
we will largely limit our discussion to that case.

As mentioned previously, the common wisdom is that $\overline{\uD
  3}$-branes break supersymmetry explicitly.  However, there are two
possible meanings to ``explicit'' breaking: either the breaking is
spontaneous but the scale of breaking is so high so that the
low-energy action effectively exhibits explicit breaking after
truncating the operators beyond a certain mass dimension (this is, for
example in the AV Lagrangian~\eqref{eq:AV_Lagrangian} where the
marginal operator alone does not exhibit supersymmetry), or the
breaking is truly explicit in that it exhibits $\cN=0$ at arbitrarily
high energies.  In the absence of warping, the scale of breaking is
naturally expected to be the compactification or string
scale\footnote{An exception to this is of course dynamical
  supersymmetry breaking in which the scale of breaking can naturally
  be much lower.} and so the distinction between explicit and
spontaneous breaking is perhaps not important.  However, for the case
of an $\overline{\uD 3}$-brane which is naturally attracted to regions
of large redshift, the scale of supersymmetry breaking may be warped
down and so the distinction may be relevant.  Before reviewing the
circumstantial evidence in this paper, let us first review some
heuristic reasoning for why the $\overline{\uD 3}$s might be expected
to break supersymmetry spontaneously.

The first is simply the statement that an $\overline{\uD 3}$-brane
represents a particular state in a supersymmetric theory, namely
string theory.  That is, whatever the fundamental description of
string theory is, it admits configurations, such as flat 10d/11d
space, that preserves 32 supercharges and therefore the theory itself
has 32 supercharges.  Any other state in the theory that preserves
fewer supercharges is still a state in a supersymmetric theory and so
those supercharges are, by definition spontaneously broken, though, as
mentioned previously, the scale of breaking may be beyond the scale at
which field theory is applicable\footnote{One possible exception to
  this is an orientifold plane which in a perturbative treatment
  literally removes fields from the spectrum.  However orientifolds,
  like D-branes, ultimately map to dynamical objects (M-branes and
  gravitational monopoles) in M-theory and so ought be treated on the
  same footing as D-branes in this sense, though the scale of breaking
  is expected to be non-perturbatively high.}.  Furthermore, as stated
previously in this work, the breaking of supersymmetry by a D-brane
should be entirely analogous to the breaking of translational symmetry
and the latter is  an example of spontaneous
breaking.  It is occasionally argued that the $\overline{\uD 3}$s in a
GKP geometry break supersymmetry explicitly because they ``project
out'' the supercharges preserved by the background.  However, they
again do so in a way that is completely analogous to the projecting
out of the translational symmetries associated to translating the
brane.  Another way of stating this argument is that the
$\overline{\uD 3}$ couples anti-holomorphically to some fields when
supersymmetry demands holomorphic couplings (e.g. the gauge kinetic
function for a $\overline{\uD 3}$-brane is proportional to
$\overline{\tau}$ rather than $\tau$).  However, the coupling is of
course holomorphic with respect to the conjugate complex structure.
That is, while the action for a $\uD 3$-brane will have actions that
are expressible as $\int\ud^{4}x\,\ud^{2}\theta\cdots$, for an
$\overline{\uD 3}$-brane, the same term will be
$\int\ud^{4}x\,\ud^{2}\theta'$ for some other fermionic coordinate
$\theta'$. This integral over a part of the $\cN_{4}=8$ superspace of
type-IIB that is different than the part integrated over by a $\uD
3$-brane is entirely analogous to the integration over a particular
part of bosonic space (namely the worldvolume) for the $\uD p$-brane
action.  That is, integrating over part of superspace is, in terms of
the breaking of supersymmetry, on the same footing as only integrating
over part of the bosonic space.  A problem very similar in spirit,
involving the spontaneous breaking of $\cN_{4}=2$ to $\cN_{4}=1$ was
considered in~\cite{Ambrosetti:2009za}.

In this work, our first line of evidence towards the spontaneous
breaking was from a stack of $\uD 3$-branes probing a
non-supersymmetric perturbation to an $\cN_{4}=1$ GKP
compactification, a system that has been considered
previously~\cite{Camara:2003ku,Grana:2003ek,Burgess:2006mn}.  A
supersymmetric GKP compactification is characterized by (among other
criteria) $G_{-}=0$ and $\Phi_{-}=0$. A small non-zero perturbation to
the latter, which is sourced ``directly'' by $\overline{\uD
  3}$-branes, perturbs the geometry and fluxes in many directions, at
least when $G_{+}\neq 0$ before the perturbation.  Among these is a
perturbation to the internal unwarped metric such that the metric is
no longer K\"ahler, at least with respect to the unperturbed
structures.  Since the internal metric is identified with the
matter-field metric for a $\uD 3$ probing the geometry, this
corresponds to a marginal deformation of the effective field theory
describing the open-string fluctuations of the $\uD 3$s.  As all soft
terms are relevant operators, naively this would imply a hard breaking
of supersymmetry.  Despite this fact, we found that when a very
natural, albeit non-holomorphic, field redefinition is performed, the
marginal operators are related by supersymmetry and thus the breaking
is soft.  Although one must be careful to not conflate soft breaking
with spontaneous breaking and hard breaking with explicit breaking,
from the Wilsonian point of view, explicit breaking is generically
expected to be hard\footnote{However, see~\cite{Dymarsky:2011ve} for
  an interesting example of a string construction in which the
  field-theory dual exhibits soft explicit breaking.} while
spontaneous breaking is soft (so long as complete multiplets remain in
the low-energy theory and even then the Lagrangian is soft only up to
some potentially hard relevant operators, which were absent for the
probe $\uD 3$s).  We thus take the non-generic non-supersymmetric
Lagrangian of the probe $\uD 3$s as an indication that breaking of
supersymmetry may be spontaneous.

From the field theory point of view, the breaking is a little unusual
in that spontaneous breaking of supersymmetry is usually accomplished
by way of some non-vanishing $F$-term or $D$-term which do not
themselves alter kinetic terms. In contrast, the probe $\uD 3$s do
experience such a deformation.  Further, the field redefinition
discussed above is non-holomorphic in the original set of fields.
This suggests that the ``least'' broken supersymmetry is not that of
the original GKP, but instead some linear combination of this
supersymmetry and others broken by GKP.  More precisely, an
$\cN_{4}=1$ GKP compactification breaks 28 of the 32 supercharges of
type-IIB.  These supercharges can be thought of as being spontaneously
broken, but since this breaking occurs at a much higher scale than
many scales of interest, we can for the most part ignore these
charges.  That is, one should in principle be able to treat the theory
as having non-linearly realized supersymmetries, but as discussed
above, this gains us very little in terms of practical value (for
example, it tells us about the higher-order terms in the fermionic
action for $\uD p$-branes, but the scale of suppression will typically
be the string scale).  When an $\overline{\uD 3}$-brane is added to a GKP
compactification, the remaining four supercharges are also broken, but
the $\cN_{4}=1$ that is most conveniently thought of as being
spontaneously broken is not quite the one preserved by GKP but instead
includes an admixture of the charges broken by GKP.  This is
reflected in both the non-holomorphic field redefinition and mixing
between the gravitino that gauges the supersymmetry preserved by GKP
and those that are lifted in GKP.  The lightest gravitino should not
be $\psi_{\mu}^{\mathrm{GKP}}$ but should instead include a linear
combination of $\psi_{\mu}^{\mathrm{GKP}}$ and the gravitini lifted by
the fluxes and curvature.

A gap in this perspective is the extension of the action for the $\uD
3$s to irrelevant order, as supersymmetry restricts more than just the
relevant and marginal operators that we considered here.  This was
reflected even in the supersymmetric case as the
Lagrangian~\eqref{eq:4d_D3_Lagrangian} exhibits $\cN_{4}=4$ through
marginal order while the irrelevant operators coming, for example,
from the non-trivial K\"ahler metric reveal it to be $\cN_{4}=1$.  In
the case of spontaneously broken $\cN_{4}=1$, supergravity imposes
that the target space metric is K\"ahler, while the internal metric
(which as stated previously is the target space metric for probe $\uD
3$s) resulting from an $\overline{\uD 3}$ appears to be generically
non-K\"ahler (at least when expressed in the complex structure of the
unperturbed geometry).  One possibility is that the backreaction is in
fact still K\"ahler, though the corresponding structures would likely
differ from~\eqref{eq:structure}.  We do not supply any evidence in
favor of this possibility which would be a very non-trivial
consequence of the supergravity equations of motion\footnote{In fact
  it is quite unlikely as there are even supersymmetric
  compactifications that are not K\"ahler.}.  Another possibility is
that the open-string moduli on the $\uD 3$ are not the correct
K\"ahler coordinates, but instead the correct coordinates are
combinations of the open-string fields on the $\uD 3$ and
closed-string fields, which occurs even in the supersymmetric case
(see, e.g.~\cite{Grana:2003ek,Jockers:2004yj}).  This would be a very
interesting case to check more precisely, but requires more work as
even the theory for closed strings alone is not wholly understood in
flux compactifications, whether or not supersymmetry is present.  Note
that since the structures~\eqref{eq:structure} are not integrable, we
have not demonstrated that the supersymmetry that is softly broken
in~\eqref{eq:final_D3_result} is globally well-defined.  It would be
important to work out whether such a globally well-defined
supersymmetry exists.

From the point of view of the $\uD 3$s and setting aside the
fluctuations of the $\overline{\uD 3}$ for a moment, it may seem
almost obvious that the breaking is spontaneous.  The DBI and CS
actions for the $\uD 3$-branes describes the interaction between the
light open strings and light closed strings and placing the $\uD 3$
brane in the $\overline{\uD 3}$ background amounts to just setting
certain expectation values for these closed string fields at a single
point in the $\uD 3$-brane moduli space.  It may be that the soft
structure of the Lagrangian is a consequence of the locality of the
$\uD 3$s.  Indeed it would be valuable to repeat this analysis for,
for example, $\uD 7$-branes or closed strings.  However, as mentioned
previously, even in supersymmetric cases the incorporation of warping
and fluxes into the effective action for strings that are not
localized at a point in the internal space can be
involved\footnote{See, for
  example~\cite{Burgess:2006mn,DeWolfe:2002nn,*Giddings:2005ff,*Frey:2006wv,*Douglas:2007tu,*Shiu:2008ry,*Douglas:2008jx,*Frey:2008xw,*Marchesano:2008rg,*Martucci:2009sf,*Camara:2009xy,*Marchesano:2010bs,*Grimm:2012rg}.},
and we leave such analyses for future work (though see, e.g.,
~\cite{Burgess:2006mn,Lust:2008zd,Benini:2009ff} for some progress in this
direction).

If the non-supersymmetric fluxes do break supersymmetry spontaneously,
then there must exist a fermion that is massless in the
$m_{\mathrm{p}}\to\infty$ limit.  The easiest case to consider is when
the fluxes result from the backreaction of an $\overline{\uD 3}$, and
we argued for the existence of such a massless fermion, namely the
gaugino on the $\overline{\uD 3}$.  If this identification is correct,
then the $\overline{\uD 3}$ brane should be thought of as $D$-term
breaking, though $F$-term breaking in the closed-string sector would
consequently result.  This possibility was also raised
in~\cite{Camara:2003ku}.  However, to make a solid case for
identification of this mode as the goldstino, there is still work to
be done.  The first, and most important, would be to demonstrate that
supersymmetry is still realized non-linearly on the anti-branes.
Here, we primarily made reference to previous work
(e.g.~\cite{Aganagic:1996nn}), but it would be worthwhile to see this
explicitly in the case at hand.  It would then be interesting to see
how the super-Higgs mechanism is realized in this setup, and to show
precisely which gravitino is lightest.  Finally, although the
discussion above focused on the case in which anti-branes were the
source of the breaking of supersymmetry, many of the points go through
for other backgrounds as well.  Indeed, the softness of the $\uD 3$
Lagrangian is a consequence of the fact that all of the marginal
operators are (to leading order in $\ell_{\us}$) controlled by the
same closed-string field, namely the internal metric, and therefore
the $\uD 3$ Lagrangian will apparently be soft in any background.  If
non-supersymmetric fluxes can always be interpreted as spontaneous
breaking, then one should be able to identify the goldstino and
understand its physics even when the fluxes do not result from the
backreaction of an anti-brane. We hope to return to these questions in
the near future.

\acknowledgments

It is our pleasure to thank David Marsh, Erik Plauschinn, and Yoske
Sumitomo for useful discussions.  We are especially indebted to Pablo
C\'amara, Liam McAllister, and Timm Wrase for their careful reading of
and comments on an early draft of this manuscript.  PM and GS further
acknowledge Michael Kiewe for collaboration on related topics.  PM
thanks the Institute for Advanced Study at Hong Kong University of
Science and Technology and the University of Wisconsin-Madison for
their hospitality during this investigation.  GS thanks the University
of Amsterdam for hospitality during the completion of this work, while
being the Johannes Diderik van der Waals Chair. PM and GS also
acknowledge the Simons Center for Geometry and Physics for
hospitality, especially during the ``String phenomenology'' workshop, April
23-27, 2012, and Uppsala University for their hospitality and
providing an environment for stimulating conversation on relevant
topics during the ``Brane backreaction, fluxes and meta-stable vacua
in string theory'' workshop, May 2-4, 2012.  The work of PM is
supported by the NSF under grant PHY-0757868.  The work of GS and FY
is supported in part by a DOE grant under contract DE-FG-02-95ER40896,
and a Cottrell Scholar Award from Research Corporation.

\appendix

\section{\label{app:conv}Conventions}

We work with the type-IIB superstring in the supergravity limit and
largely follow the conventions of~\cite{Martucci:2005rb}.  In the 10d
Einstein frame, the bosonic pseudo-action is
\begin{subequations}
\label{eq:IIB_action}
\begin{align}
  S_{\mathrm{IIB}}=&S_{\mathrm{IIB}}^{\mathrm{NS}}
  +S_{\mathrm{IIB}}^{\mathrm{R}}
  +S_{\mathrm{IIB}}^{\mathrm{CS}},\\
  S_{\mathrm{IIB}}^{\mathrm{NS}}=&
  \frac{1}{2\ka^{2}_{10}}\int\ud^{10}x\,
  \sqrt{-\det\left(\hat{g}\right)}
  \biggl[\hat{R}-\frac{1}{2}\hat{g}^{MN}\partial_{M}\phi\partial_{N}\phi
  -\frac{1}{2}\ue^{-\phi}\hat{H}_{\left(3\right)}^{2}
  \biggr],\\
  S_{\mathrm{IIB}}^{\mathrm{R}}=&
  -\frac{1}{4\ka^{2}_{10}}
  \int\ud^{10}x\,
  \sqrt{-\det\left(\hat{g}\right)}
  \biggl[\ue^{2\phi}\hat{F}_{\left(1\right)}^{2}
  +\ue^{\phi}\hat{F}_{\left(3\right)}^{2}
  +\frac{1}{2}\hat{F}_{\left(5\right)}^{2}\biggr],\\
  S_{\mathrm{IIB}}^{\mathrm{CS}}=&
  \frac{1}{4\ka^{2}_{10}}
  \int C_{\left(4\right)}\wedge H_{\left(3\right)}\wedge F_{\left(3\right)},
\end{align}
\end{subequations}
in which the 10d gravitational constant is
$2\ka_{10}^{2}=\frac{1}{2\pi}\ell_{\us}^{8}g_{\us}^{2}$ where
$\ell_{\us}=2\pi\sqrt{\alpha'}$ is the string length.  $\hat{R}$ is
the Ricci scalar built from the 10d Einstein-frame metric
$\hat{g}_{MN}$ which is related to the 10d string-frame metric by
$\hat{g}_{MN}=\ue^{-\phi/2}\hat{g}_{MN}^{\left(\mathrm{s}\right)}$.  $\phi$ is
the dilaton defined so that the string coupling is
$g_{\us}\ue^{\phi}$.  The NS-NS $2$-form potential is
$B_{\left(2\right)}$ and the R-R potentials are $C_{\left(p\right)}$
for $p=0,2,4$.  The gauge-invariant field strengths are
\begin{equation}
  H_{\left(3\right)}=\ud B_{\left(2\right)},\quad
  F_{\left(1\right)}=\ud C_{\left(0\right)},\quad
  F_{\left(3\right)}=\ud C_{\left(2\right)}
  +C_{\left(0\right)}\wedge H_{\left(3\right)},\quad
  F_{\left(5\right)}=\ud C_{\left(4\right)}
  +C_{\left(2\right)}\wedge H_{\left(3\right)}.
\end{equation}
$F_{\left(5\right)}$ is constrained at the level of the equations of
motion to satisfy the self-duality constraint
$F_{\left(5\right)}=\hat{\ast}F_{\left(5\right)}$ in which
$\hat{\ast}$ is the 10d Hodge-$\ast$, $\bigl(\hat{\ast}
F\bigr)_{MNPQR}=\frac{1}{5!}\hat{\ep}_{MNPQR}^{\phantom{MNPQR}STLKI}F_{STLKI}$. We
use the convention that in flat space $\hat{\ep}_{01\cdots 9}=+1$.  For a
$p$-form we define
\begin{equation}
  \hat{\Omega}_{\left(p\right)}^{2}=\frac{1}{p!}
  \hat{g}^{M_{1}N_{1}}\cdots\hat{g}^{M_{p}N_{p}}
  \Omega_{M_{1}\cdots M_{p}}\Omega_{N_{1}\cdots N_{p}}.
\end{equation}
More generally, $\hat{\phantom{a}}$ will indicate objects pertaining
to the 10d metric $\hat{g}_{MN}$.

The equations of motion that follow from~\eqref{eq:IIB_action} are
(see, e.g.~\cite{Polchinski:2000uf})
\begin{subequations}
\label{eq:IIB_eom}
\begin{align}
  0=&\hat{R}_{MN}-\frac{1}{2}\partial_{M}\phi\partial_{N}\phi
  -\frac{1}{2}\ue^{2\phi}F_{M}F_{N}
  -\frac{1}{2\cdot 2!}\ue^{-\phi}{H}_{MPQ}\hat{H}_{N}^{\phantom{N}PQ}
  -\frac{1}{2\cdot 2!}\ue^{\phi}F_{MPQ}\hat{F}_{N}^{\phantom{N}PQ}
  \notag\\
  &-\frac{1}{4\cdot 4!}{F}_{MPQRS}\hat{F}_{N}^{\phantom{N}PQRS}
  +\frac{1}{8}\hat{g}_{MN}\biggl[
  \ue^{-\phi}\hat{H}_{\left(3\right)}^{2}
  +\ue^{\phi}\hat{F}_{\left(3\right)}^{2}\biggr],\\
  0=&\hat{\nabla}^{2}\phi-\ue^{2\phi}\hat{F}_{\left(1\right)}^{2}
  -\frac{1}{2}\ue^{\phi}\biggl[\hat{F}_{\left(3\right)}^{2}-
  \ue^{-2\phi}\hat{H}_{\left(3\right)}^{2}\biggr],\\
  0=&\hat{\nabla}^{M}\bigl(\ue^{2\phi}F_{M}\bigr)
  -\frac{\ue^{\phi}}{3!}H_{MNP}\hat{F}^{MNP},\\
  0=&\ud\hat{\ast}\bigl(\ue^{-\phi}H_{\left(3\right)}+C_{\left(0\right)}\ue^{\phi}
  F_{\left(3\right)}\bigr)-F_{\left(5\right)}\wedge F_{\left(3\right)},\\
  0=&\ud\hat{\ast}\bigl(\ue^{\phi}F_{\left(3\right)}\bigr)+F_{\left(5\right)}\wedge
  H_{\left(3\right)},\\
  0=&\ud\hat{\ast} F_{\left(5\right)}+H_{\left(3\right)}\wedge F_{\left(3\right)}.
\end{align}
Here, $\hat{R}_{MN}$ is the Ricci tensor and we have imposed
self-duality on $F_{\left(5\right)}$.  In addition, we have the
Bianchi identities
\begin{equation}
  \ud H_{\left(3\right)}=0,\quad \ud F_{\left(1\right)}=0,\quad
  \ud F_{\left(3\right)}=F_{\left(1\right)}\wedge H_{\left(3\right)},\quad
  \ud F_{\left(5\right)}=F_{\left(3\right)}\wedge H_{\left(3\right)}.
\end{equation}
\end{subequations}  

Along with these bosonic modes, type-IIB supergravity contains a pair
of 32-component Majorana-Weyl dilatini $\hat{\chi}^{1,2}$ and a pair
of Majorana-Weyl-Rarita-Schwinger gravitini $\hat{\Psi}_{M}^{1,2}$.
We take these modes to be right-handed in the sense that
$\Ga_{\left(10\right)}\hat{\Psi}_{M}^{i}=\hat{\Psi}_{M}^{i}$ where the
10d-chirality operator $\Ga_{\left(10\right)}$ is defined
by~\eqref{eq:chirality_op}.  These can be used to construct so-called
double spinors
\begin{equation}
  \hat{\chi}=\begin{pmatrix}\hat{\chi}^{1}\\\chi^{2}\end{pmatrix},\quad
  \hat{\Psi}^{M}=\begin{pmatrix}\hat{\Psi}_{M}^{1}\\
    \hat{\Psi}_{M}^{2}\end{pmatrix}.
\end{equation}
The action for the closed-strings fermions will not be used be used
here, but the combined action is invariant under
$\cN_{10}=\left(2,0\right)$ supersymmetry under which the fermions
transform as
\begin{equation}
  \delta_{\hat{\ep}}\hat{\chi}=\cO\hat{\ep},\quad
  \delta_{\hat{\ep}}\hat{\Psi}_{M}=\hat{\cD}_{M}\hat{\ep},
\end{equation}
in which $\hat{\ep}$ is a double right-handed Majorana-Weyl spinor and
\begin{subequations}
\label{eq:susy_vars}
\begin{align}
  \hat{\cO}=&\frac{1}{2}\hat{\slashed{\partial}}\phi
  -\frac{1}{2}e^{\phi}\hat{\slashed{F}}_{\left(1\right)}
  \ui\sig^{2}
  -\frac{1}{4}\ue^{\phi/2}\cG^{-},\\
  \hat{\cD}_{M}=&\hat{\nabla}_{M}
  +\frac{1}{4}\ue^{\phi}\partial_{M}C_{\left(0\right)}
  \ui\sig^{2}
  +\frac{1}{8}\ue^{\phi/2}
  \bigl(\cG^{+}\hat{\Ga}_{M}+\frac{1}{2}\hat{\Ga}_{M}\cG^{+}\bigr)
  +\frac{1}{16}\hat{\slashed{F}}_{\left(5\right)}\hat{\Ga}_{M}\ui\sig^{2},
\end{align}
\end{subequations}
in which
\begin{equation}
\label{eq:curly_G_2}
  \cG^{\pm}=\hat{\slashed{F}}_{\left(3\right)}\sig^{1}\pm 
  \ue^{-\phi}\hat{\slashed{H}}_{\left(3\right)}\sig^{3}.
\end{equation}
For a $p$-form,
\begin{equation}
  \hat{\slashed{\Omega}}_{\left(p\right)}
  :=\frac{1}{p!}\Omega_{M_{1}\cdots M_{p}}\hat{\Ga}^{M_{1}\cdots M_{p}},
\end{equation}
in which
\begin{equation}
  \hat{\Ga}^{M_{1}\cdots M_{p}}=
  \hat{\Ga}^{\left[M_{1}\right.}\cdots\hat{\Ga}^{\left. M_{p}\right]},
\end{equation}
where $\left[\cdots\right]$ denotes averaging over signed
permutations, e.g.,
\begin{equation}
  X^{\left(MPQ\right)}=\frac{1}{3!}\biggl(X^{MPQ}+X^{MQP}+\cdots\biggr),\quad
  X^{\left[MPQ\right]}=\frac{1}{3!}\biggl(X^{MPQ}-X^{MQP}+\cdots\biggr).
\end{equation}
Unless otherwise noted, $\Ga$-matrices on double spinors as
\begin{equation}
  \hat{\Ga}_{M}\hat{\ep}=\begin{pmatrix}\hat{\Ga}_{M}\hat{\ep}^{1}\\ 
    \hat{\Ga}_{M}\hat{\ep}^{2}\end{pmatrix}.
\end{equation}
The Pauli matrices appearing in~\eqref{eq:susy_vars} act on the
so-called extension space.  For example,
\begin{equation}
  \sig^{1}\begin{pmatrix}\hat{\ep}^{1}\\\hat{\ep}^{2}\end{pmatrix}
  =\begin{pmatrix}\hat{\ep}^{2}\\\hat{\ep}^{1}\end{pmatrix}.
\end{equation}

$\hat{\nabla}$ is the covariant derivative defined by
\begin{subequations}
\label{eq:spin_connection}
\begin{equation}
  \hat{\nabla}_{M}=\partial_{M}+\frac{1}{4}\hat{\omega}_{M}^{\phantom{M}\ul{NP}}
  \hat{\Ga}_{\ul{NP}},
\end{equation}
where $\hat{\omega}_{M}^{\phantom{M}\ul{NP}}$ are the components of the spin
connection
\begin{equation}
  \hat{\omega}_{M}^{\phantom{M}\ul{NP}}=\frac{1}{2}\hat{e}_{M}^{\phantom{M}\ul{Q}}
  \bigl(\hat{T}_{\ul{Q}}^{\phantom{\ul{Q}}\ul{N}\ul{P}}
  -\hat{T}^{\ul{N}\ul{P}}_{\phantom{\ul{N}\ul{P}}\ul{Q}}-
  \hat{T}^{\ul{P}\phantom{\ul{Q}}\ul{N}}_{\phantom{\ul{P}}\ul{Q}}\bigr),\quad
  \hat{T}^{\ul{M}}_{\phantom{\ul{M}}\ul{N}\ul{P}}=
  \bigl(\hat{e}^{Q}_{\phantom{Q}\ul{N}}\hat{e}^{R}_{\phantom{R}\ul{P}}-
  \hat{e}^{Q}_{\phantom{Q}\ul{P}}\hat{e}^{R}_{\phantom{R}\ul{N}}\bigr)
  \partial_{R}\hat{e}_{Q}^{\phantom{Q}\ul{M}},
\end{equation}
\end{subequations}
in which $\hat{e}_{M}^{\phantom{M}\ul{N}}$ are the vielbein defining
the local frame $\hat{e}^{\ul{N}}=\hat{e}_{M}^{\phantom{M}\ul{N}}\ud
x^{M}$ and $\hat{e}^{M}_{\phantom{M}\ul{N}}$ are the inverse vielbein.

For the $\hat{\Ga}$-matrices, we choose a basis that is useful for the
decomposition $\SO{9,1}\to\SO{3,1}\times\SO{6}$.  In $3+1$ dimensions,
we take in a local frame
\begin{equation}
  \ga^{\ul{\mu}}=\begin{pmatrix}
    0 & -\bar{\sig}^{\ul{\mu}} \\ \sig^{\ul{\mu}} & 0
  \end{pmatrix},
\end{equation}
in which
\begin{equation}
  \sig^{\ul{\mu}}=\bigl(1,\bs{\sig}\bigr),\quad
  \bar{\sig}^{\ul{\mu}}=\bigl(1,-\bs{\sig}\bigr).
\end{equation}
where $\boldsymbol{\sig}$ are the usual Pauli matrices
\begin{equation}
  \sig^{1}=\begin{pmatrix}0 & 1 \\ 1 & 0\end{pmatrix},\quad
  \sig^{2}=\begin{pmatrix}0 & -\ui \\ \ui & 0\end{pmatrix},\quad
  \sig^{3}=\begin{pmatrix}1 & 0 \\ 0 & -1\end{pmatrix}.
\end{equation}
The $\ga$-matrices then satisfy
\begin{equation}
  \bigl\{\ga^{\ul{\mu}},\ga^{\ul{\nu}}\bigr\}=2\eta^{\ul{\mu}\ul{\nu}}.
\end{equation}
The 4d chirality operator is
\begin{equation}
  \ga_{\left(4\right)}=\frac{\ui}{4!} \ve_{\mu_{1}\cdots\mu_{4}}
  \ga^{\mu_{1}\cdots\mu_{4}}
  =\begin{pmatrix}
    \II_{2} & 0 \\ 0 & -\II_{2}
  \end{pmatrix},
\end{equation}
in which $\ve_{0123}=+\sqrt{\det\left(g_{\mu\nu}\right)}$.  Since
$\ga^{\ul{2}}$ is the only imaginary $\ga$-matrix, the 4d Majorana matrix
is
\begin{equation}
  B_{4}=\ga_{\left(4\right)}\ga^{\ul{2}}=
  \begin{pmatrix} 0 & \sig^{2} \\ -\sig^{2} & 0
  \end{pmatrix},
\end{equation}
and satisfies $B_{4}B_{4}^{\ast}=\II_{4}$,
$\ga^{\mu}B_{4}=B_{4}\ga^{\mu\ast}$, and
$\ga_{\left(4\right)}B_{4}=-B_{4}\ga_{\left(4\right)}^{\ast}$.

We will make use the dotted, undotted notation
of~\cite{Wess:1992cp} and write a 4d Dirac spinor as
\begin{equation}
  \psi=\begin{pmatrix}
    \ui\bar{\psi}_{\uR}^{\dot{\alpha}} \\ \psi_{\uL\alpha}
  \end{pmatrix},
\end{equation}
where we raise and lower indices with $\ep^{12}=\ep_{21}=1$.

In 6 dimensions, we take in an orthonormal frame\footnote{In the main
  text, we drop the $\tilde{\phantom{a}}$ appearing above these
  $\ga$-matrices since context should hopefully make clear whether we
  mean $\SO{3,1}$ or $\SO{6}$ $\ga$-matrices.}
\begin{align}
  \tilde{\ga}^{1}=&\sig^{1}\otimes\II_{2}\otimes\II_{2},&
  \tilde{\ga}^{4}=&\sig^{2}\otimes\II_{2}\otimes\II_{2},\notag\\
  \tilde{\ga}^{2}=&\sig^{3}\otimes\sig^{1}\otimes\II_{2},&
  \tilde{\ga}^{5}=&\sig^{3}\otimes\sig^{2}\otimes\II_{2},\\
  \tilde{\ga}^{3}=&\sig^{3}\otimes\sig^{3}\otimes\sig^{1},&
  \tilde{\ga}^{6}=&\sig^{3}\otimes\sig^{3}\otimes\sig^{2}.\notag
\end{align}
They satisfy
\begin{equation}
  \bigl\{\tilde{\ga}^{\ul{m}},\tilde{\ga}^{\ul{n}}\bigr\}=2\delta^{\ul{m}\ul{n}}.
\end{equation}
The 6d chirality operator is then
\begin{equation}
  \tilde{\ga}_{\left(6\right)}
  =-\frac{\ui}{6!}{\tilde{\ve}}_{m_{1}\cdots m_{6}}{\tilde{\ga}}^{m_{1}\cdots
    m_{6}}=\sig_{3}\otimes\sig_{3}\otimes\sig_{3}.
\end{equation}
The 6d Majorana matrix is
\begin{equation}
  \tilde{B}_{6}=\tilde{\ga}^{4}\tilde{\ga}^{5}\tilde{\ga}^{6}
  =\ui \sig^{2}\otimes\sig^{1}\otimes\sig^{2}.
\end{equation}
It satisfies $\tilde{B}_{6}\tilde{B}_{6}^{\ast}=\II_{8}$,
$\tilde{\ga}^{m}\tilde{B}_{6}=-\tilde{B}_{6}\tilde{\ga}^{m\ast}$, and
$\tilde{\ga}_{\left(6\right)}\tilde{B}_{6}=-\tilde{B}_{6}\tilde{\ga}_{\left(6\right)}^{\ast}$.

From these, we define the 10d $\Ga$-matrices,
\begin{equation}
\label{eq:10d_gamma_matrices}
  \hat{\Ga}^{\ul{\mu}}=\ga^{\ul{\mu}}\otimes\II_{8},\quad
  \hat{\Ga}^{\ul{m}}=\ga_{\left(4\right)}\otimes\tilde{\ga}^{\ul{m}},
\end{equation}
where the second equality should be read as
$\hat{\Ga}^{4}=\ga_{\left(4\right)}\otimes\tilde{\ga}^{1}$, etc.  They satisfy
\begin{equation}
  \bigl\{\hat{\Ga}^{\ul{M}},\hat{\Ga}^{\ul{N}}\bigr\}=2\hat{\eta}^{\ul{M}\ul{N}}.
\end{equation}
The 10d chirality operator is then
\begin{equation}
  \label{eq:chirality_op}
  \Ga_{\left(10\right)}
  =\frac{1}{10!}\hat{\ve}_{M_{1}\cdots M_{10}}\hat{\Ga}^{M_{1}\cdots M_{10}}
  =\ga_{\left(4\right)}\otimes\tilde{\ga}_{\left(6\right)}.
\end{equation}
The 10d Majorana matrix
\begin{equation}
  \hat{B}_{10}=\hat{\Ga}^{2}\hat{\Ga}^{7}\hat{\Ga}^{8}\hat{\Ga}^{9}
  =-B_{4}\otimes\tilde{B}_{6},
\end{equation}
satisfies $\hat{B}_{10}\hat{B}_{10}^{\ast}=\II_{32}$,
$\hat{\Ga}^{M}\hat{B}_{10}=\hat{B}_{10}\hat{\Ga}^{\ast}_{M}$, and
$\Ga_{\left(10\right)}\hat{B}_{10}=\hat{B}_{10}{\Ga}^{\ast}_{\left(10\right)}$.

\bibliography{msy}

\end{document}